\documentclass[twocolumn, twoside,journal]{IEEEtran}
\usepackage{blindtext}
\usepackage{graphicx}

\usepackage[T1]{fontenc}

\usepackage[noadjust]{cite}
\usepackage[cmex10]{amsmath}
\usepackage{hyperref}
\usepackage[mathscr]{euscript}

\usepackage{algorithm}
\usepackage{algorithmic}
\usepackage{enumitem}
\usepackage{subcaption}
\usepackage{cleveref}

\usepackage{color}
\usepackage[table]{xcolor}

\usepackage{amssymb}



\usepackage{caption}

\usepackage{textcomp}
\usepackage{bbold}

\hypersetup{
	colorlinks,
	citecolor=black,
	filecolor=black,
	linkcolor=black,
	urlcolor=black
}
\newtheorem{theorem}{Theorem}
\newtheorem{lemma}{Lemma}
\newtheorem{remark}{Remark}
\newtheorem{definition}{Definition}
\newtheorem{proposition}{Proposition}
\newtheorem{corollary}{Corollary}
\newtheorem{assumption}{Assumption}
\let\oldtheorem\theorem
\let\endoldtheorem\endtheorem
\def\theorem{\begingroup \oldtheorem }
\def\endtheorem{ \hfill $\square$\endoldtheorem \endgroup}

\let\oldlemma\lemma
\let\endoldlemma\endlemma
\def\lemma{\begingroup \oldlemma }
\def\endlemma{ \hfill $\square$\endoldlemma \endgroup}

\let\oldremark\remark
\let\endoldremark\endremark
\def\remark{\begingroup \oldremark }
\def\endremark{ \hfill $\square$\endoldremark\endgroup}

\let\oldproposition\proposition
\let\endoldproposition\endproposition
\def\proposition{\begingroup \oldproposition}
\def\endproposition{ \hfill $\square$\endoldproposition\endgroup}

\let\oldcorollary\corollary 
\let\endoldcorollary\endcorollary
\def\corollary{\begingroup \oldcorollary}
\def\endcorollary{ \hfill $\square$\endoldcorollary\endgroup}

\let\olddefinition\definition
\let\oldenddefinition\enddefinition
\def\definition{\begingroup \olddefinition  }
\def\enddefinition{ \hfill $\square$\oldenddefinition\endgroup}

\let\oldassumption\assumption
\let\oldendassumption\endassumption
\def\assumption{\begingroup \oldassumption  }
\def\endassumption{ \hfill $\square$\oldendassumption\endgroup}

\newcolumntype{P}[1]{>{\centering\arraybackslash}m{#1}}
\let\oldIEEEkeywords\IEEEkeywords
\let\oldendIEEEkeywords\endIEEEkeywords
\def\IEEEkeywords{\begingroup \oldIEEEkeywords \begingroup\normalfont\bfseries}
\def\endIEEEkeywords{ \endgroup\oldendIEEEkeywords \endgroup}
\usepackage{tabulary}
\usepackage{booktabs}

\usepackage{lettrine}

\usepackage{cite}

\usepackage{soul}

\begin{document}

\title{Generalized Wardrop Equilibrium for Charging Station Selection and Route Choice of Electric Vehicles in Joint Power Distribution and Transportation Networks}
\author{Babak~Ghaffarzadeh Bakhshayesh, Hamed~Kebriaei*,~\IEEEmembership{Senior Member,~IEEE}
	\thanks{ $^*$Corresponding Author (Hamed Kebriaei)}
		\thanks{This work was in part supported by a grant from the Institute for Research
			in Fundamental Sciences (IPM) under grant number: CS 1400-4-451.}
	\thanks{B. Ghaffarzadeh Bakhshayesh, H. Kebriaei are with the School of Electrical
		and Computer Engineering, College of Engineering, University of Tehran, Tehran, Iran. H. Kebriaei is also with the School of
		Computer Science, Institute for Research in Fundamental Sciences (IPM).}
}
\maketitle
\begin{abstract}
{
	
	This paper presents the equilibrium analysis of a game composed of heterogeneous electric vehicles (EVs) and a power distribution system operator (DSO) as the players, and charging station operators (CSOs) and a transportation network operator (TNO) as coordinators. Each EV tries to pick a charging station as its destination and a route to get there at the same time. However, the traffic and electrical load congestion on the roads and charging stations lead to the interdependencies between the optimal decisions of EVs. CSOs and the TNO need to apply some tolling to control such congestion. On the other hand, the pricing at charging stations depends on real-time distributional locational marginal pricing, which is determined by the DSO after solving the optimal power flow over the power distribution network. This paper also takes into account the local and the coupling/{infrastructure} constraints of EVs, transportation and distribution networks.
This problem is modeled as a generalized aggregative game, and then a decentralized learning method is proposed to obtain an equilibrium point of the game, which is known as variational generalized Wardrop equilibrium. The existence of such an equilibrium point and the convergence of the proposed algorithm to it are proven.
We undertake numerical studies on the Savannah city model and the IEEE 33-bus distribution network and investigate the impact of various characteristics on demand and prices.
}

\end{abstract}
\section{Introduction}
{In recent years, the increasing proliferation of electric vehicles (EVs) in transportation networks and the adoption of distributed energy sources in distribution networks have restructured and tightened the connection between these two types of networks, potentially increasing the complexity of managing both systems, simultaneously \cite{bibra2021global}. While these modifications address climate change and net-zero emission \cite{davis2018net}, uncoordinated management of large numbers of EVs in cities can have a detrimental effect on the distribution network due to transformer overloads, power flow congestion, and local energy prices \cite{muratori2018impact}. Additionally, this might result in higher congestion on both roads and charging stations as price-sensitive users seek out cheaper electricity \cite{18}. Meanwhile, customers now have real-time access to information about road traffic and power costs, thanks to the rapid growth of supporting technologies like Google Maps and PlugShare. Thus, coordinators in a smart city should leverage these new coordination tools to mitigate such negative repercussions and incentivize selfish EVs appropriately, ensuring that the rising charging demand is met through the integration of sustainable energy supplies in a socio-technical system \cite{ratliff2019perspective}.
	 In this paper, EVs are modeled as self-interested agents that want to find both the best charging station and the best way to get there. They do this by taking into account travel time, electricity price, station crowding, and personal preferences, as well as local constraints and the constraints imposed by the transportation and distribution networks. Furthermore, the distribution network operator aims to minimize the operating cost of distributed generators (DGs) while incentivizing EVs to comply with coupling constraints such as power system capacity and active power balancing constraints, which also depend on EVs' decisions. This aim can be achieved using market clearing of the power distribution system operator (DSO) based on real-time distributional locational marginal pricing (DLMP). Furthermore, charging station operators (CSOs) and the transportation network operator (TNO) impose and broadcast real-time tolls to ensure that the capacity of stations and roads {will} not be violated. It is worth mentioning that our proposed framework can be extended to incorporate other scenarios, such as competition among charging stations or information asymmetry in EVs. Therefore, considering these aspects, a decentralized method of gradient-based learning of generalized Wardrop equilibrium in a holistic model with infrastructure and coupling constraints among users is proposed.\\
	The main contributions of the proposed decentralized game-theoretical interaction of different types of users in coupled transportation and distribution networks are as follows: {
	\begin{itemize}
		\item We present a novel framework for simultaneous routing and charging station selection of selfish EVs by considering the effects of power distribution and transportation networks coordinated by the DSO, CSOs, and the TNO.
		\item The \textit{LinDistFlow} model for distribution power flow is considered to ensure that distribution network constraints are met by imposing DLMPs on the charging stations based on Lagrangian relaxation of the generalized game. Further, we introduce congestion pricing based on the average utilization of heterogeneous stations to manage stations' crowdedness. 
		\item We formalize the corresponding interaction as a generalized aggregative game and solve it by exploiting the inertial forward-reflected-backward (I-FoRB) splitting algorithm for non-strictly monotone games, ensuring the scalability and privacy of all entities. Then the existence of variational generalized Wardrop equilibrium (v-GWE) and convergence of the decentralized algorithm to a v-GWE is proved.
		\item We provide a numerical study of the Savannah city transportation network model and the IEEE 33-bus radial distribution network to explore the effects of DGs and distribution network constraints on system equilibrium as well as users' preferences.
	\end{itemize}}}
The rest of the paper is arranged as follows: We conduct a literature review in section \ref{litrev}. Then, in Section \ref{problem_formulation}, we provide the coupled model of user interactions in the transportation and power networks, followed by the decoupled model of the game. Section \ref{Decentralized} then establishes the existence of a generalized Wardrop equilibrium and proposes a decentralized algorithm which converges to a Wardrop equilibrium. Section \ref{Simulation} investigates a case study, while Section \ref{Conclusion} concludes.
\section{Literature review}\label{litrev}
Recently, there has been considerable interest in {jointly} analyzing the system-level impact of the transportation and power networks. We can categorize them into two groups. The first category includes research studies that use centralized optimization methodologies to maximize the social welfare of coupled power and transportation networks. In \cite{3}, using Lagrangian relaxations of the related social optimum optimization problem, collaborative pricing based on Locational Marginal Pricing (LMP) and between power and transportation networks is suggested to compute the social optimum of systems. It is also inferred that using marginal pricing of road flows, selfish users would follow the calculated socially optimum flow.
The authors of \cite{5} apply the interdependence model to solve the unit commitment problem and propose mixed-integer linear programming to schedule EV fleets with variable wind power generation.
The authors of \cite{10, 11} conduct a series of works on the interaction between emerging AMoD systems and power network, in which the fleet of electric self-driving vehicles servicing the on-demand transportation network is controlled centrally in collaboration with the power system.\\ 
The second group includes research papers that assume users are selfish agents who want to decide to minimize their desired cost strategically. Using a combined distribution and assignment (CDA) model of user equilibrium, \cite{1} investigated optimal pricing to compute social welfare over joint transportation and power networks. However, the CDA model's parameters must be empirically calibrated.
In \cite{2}, the fixed point of power network decision and transportation network equilibrium is attained by iteratively solving the best response of each transportation and power network. In \cite{6}, the socially optimal equilibrium of connected transportation and power networks is achieved using mixed-integer second-order cone programming and the assumption that EVs can be charged wirelessly, eliminating the need for them to drive to charging points. The authors of \cite{7} studied a bi-level model and used reinforcement learning to solve socially optimal charging pricing while mitigating uncertainty in consumers' origin-destination demand and wind power generation. A recent study \cite{12} presents a tri-level optimization technique for modeling interactions among EVs, charging stations, and electricity networks, using an iterative bi-level algorithm and simulated annealing.
The comprehensive literature review on various models of this subject can be found in \cite{13, zardini2022analysis}.\\
Although the above studies contribute to managing EVs, the following challenges remain. First, in \cite{1,2,3,5,6,7,10,11,12} and \cite{8, 9, tucker2019online, 9390363}, the system condition is computed centrally, which means either the EVs implement central coordinator suggestions in optimization formulations or the central operator finds the users' equilibrium based on complete user information in game-theoretic formulations. But perfect knowledge of users' preferences is unachievable and ignores privacy. 
The survey paper \cite{113} comprehensively reviews recent decentralized coordination mechanisms and their advantages in comparison with centralized approaches in the literature as well.\\
{
	Second, the user equilibrium models in and \cite{1,2,3,5,6,7,10,11,12} and \cite{9,tucker2019online, 9390363, 8} are based on the assumption that all users have the same cost function and constraints, i.e., the same monetary value of time and EV compatibility with charging stations. Using Beckman's potential function, traffic assignment problems result in time value homogeneity among users in studies \cite{1,2,3,5,6,7,10,11,12} and \cite{8, 9, tucker2019online, 9390363} because the hessian matrix of the potential game has to be symmetric. However, our monotone game formulation allows for the heterogeneity of users' time values. Furthermore, \cite{1,2,3,5,6,7,10,11,12} and \cite{8, 9, tucker2019online, 9390363} assume that all EVs are compatible with all charging stations.}
In \cite{3}, although marginal tolling of a transportation network operator will ensure that the users follow a socially optimal solution calculated by the central operator, it is dependent on two strong assumptions: the homogeneity of users and the central operator's complete knowledge. Moreover, the authors do not consider the algorithmic aspects of edge pricing and the unfortunate possibility that traffic routing may only be efficient with high tolls \cite{4}. So, this consideration may not be suitable in practice.
Third, as discussed in \cite{18}, even though user equilibrium is a convex problem in studies \cite{3,5,10,11}, \cite{2,6,7,12}, and \cite{8, 9} enumerating all feasible paths is {an} NP-hard problem. In contrast, the edge formulation proposed in \cite{18} eliminates the necessity of enumerating feasible paths. {The table of notations used in the paper is provided in Table \ref{notation}.
\begin{table}[htbp]\caption{Table of Notations used in the article}\label{notation}
	\centering 
	\begin{tabular}{r c p{5cm} }
		\toprule
		$\mathbb{R}$ & $\triangleq$ & set of real numbers\\
		$\mathbb{R}_{\ge 0}$ & $\triangleq$ & set of non-negative real numbers\\
		col$({{x}_{1}},\cdots,{{x}_{M}})$ & $\triangleq$ & $[{{x}_{i}}]_{i=1}^{M} = {{[x_{1}^{\top},\ldots,x_{M}^{\top}]}^{\top}}$\\
		$\text{pro}{{\text{j}}_{\mathcal{X}}}(x):{{\mathbb{R}}^{n}}\to \mathcal{X}$ & $\triangleq$ & $\arg\min_{y\in{\cal X}}||y-x||$\\
		${{\Pi}_{{{\alpha }_{i}}{{g}_{i}}+{{\iota }_{{{\mathcal{X}}_{i}}}}}}(y)$ & $\triangleq$ & $\arg\min_{z\in {{\mathcal{X}}_{i}}}{{g}_{i}}(z)+\frac{1}{2{{\alpha }_{i}}}{{\left\| z-y \right\|}^{2}}$\\
		${{\nabla }_{x}}f\left( x \right)\in {{\mathbb{R}}^{d\times n}}$ & $\triangleq$ & gradient of function $f\left( x \right):{{\mathbb{R}}^{n}}\to {{\mathbb{R}}^{d}}$\\  
		${{\left[ A \right]}_{i}}$ & $\triangleq$ & the element on row $i$ of $A$\\
		\bottomrule
	\end{tabular}
\end{table}}
\label{system_model_section}
\section{System model} \label{problem_formulation}
\begin{figure}[!ht]
	\centering
	\includegraphics[width=0.9\linewidth]{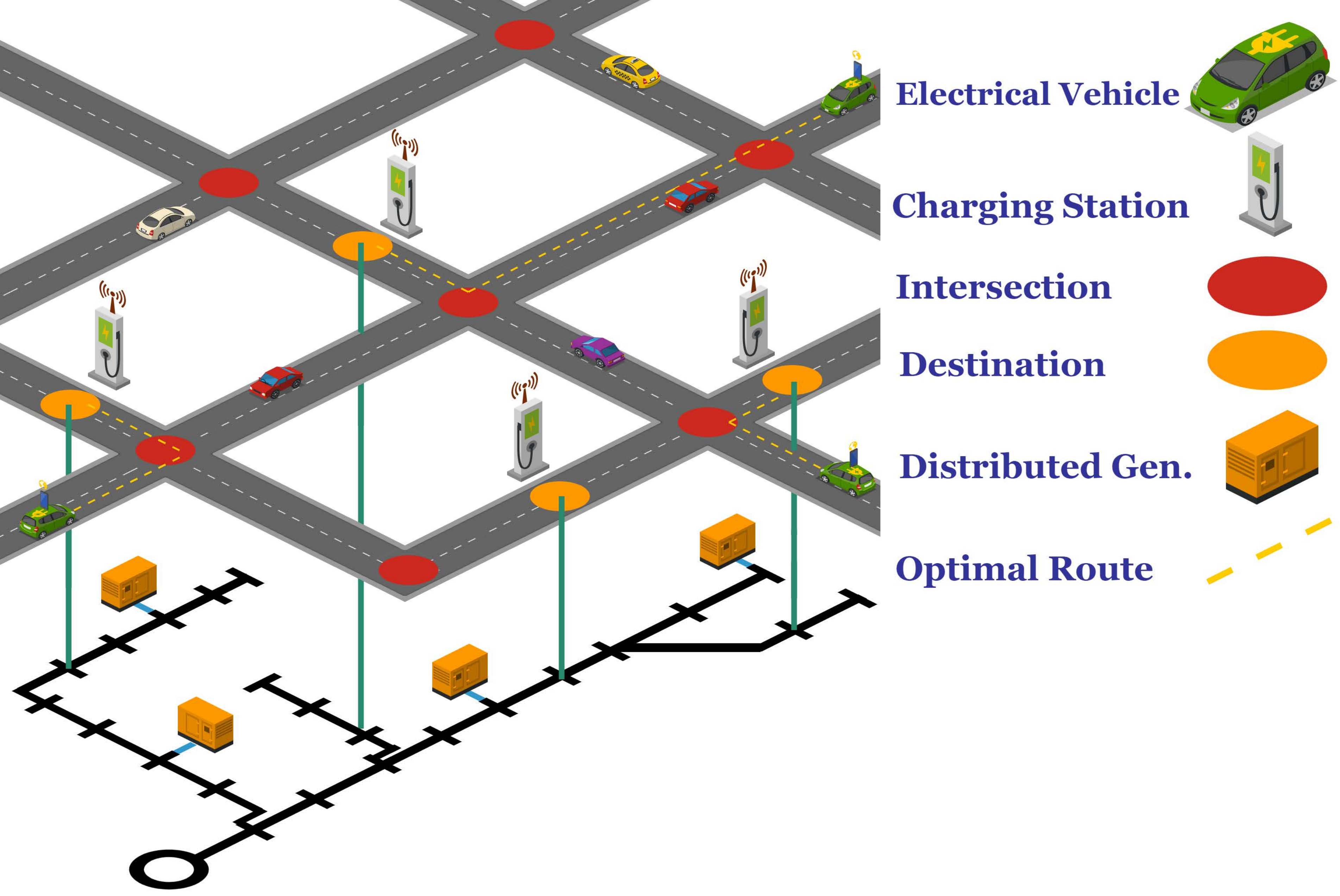}
	\caption{Transportation network, charging stations' location, power distribution network, and EVs.}
	\label{Map_power}
\end{figure}

\begin{figure}[!ht]
	\centering
	\includegraphics[width=1\linewidth]{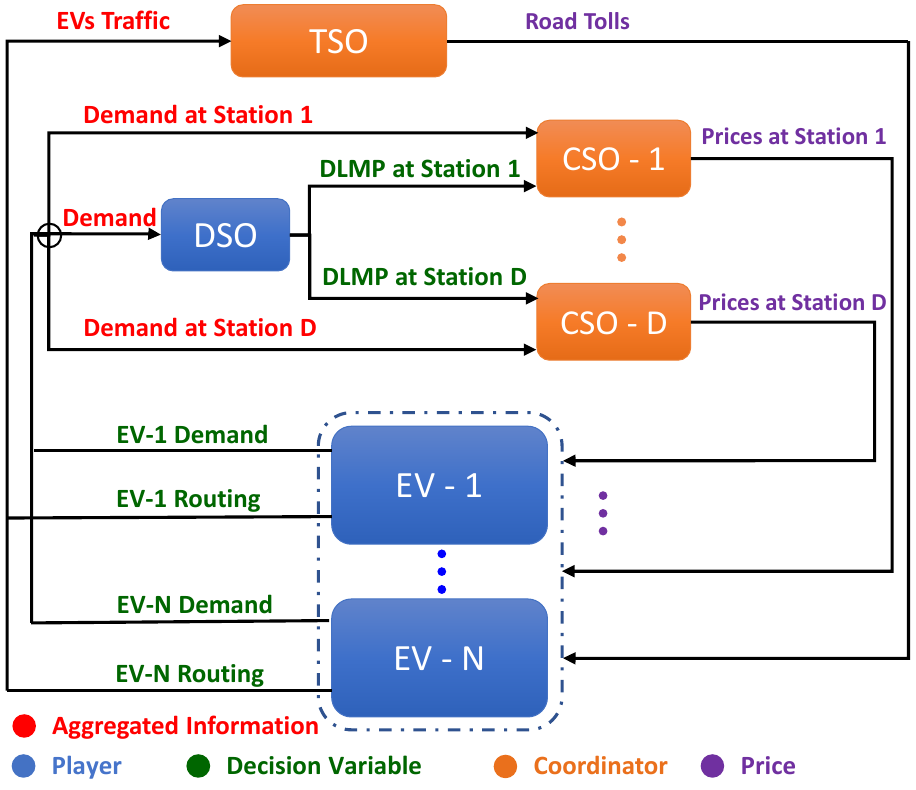}
	\caption{The information flows of the system.}
	\label{Flow}
\end{figure}

We consider a transportation network connected to an $N$-bus radial electrical distribution network at some charging stations. The networks are modeled via directed graphs ${{G}_{T}}=\left( \mathcal{V},\mathcal{E} \right)$ and ${{G}_{P}}=\left( \mathcal{N},\mathcal{L} \right)$, respectively. Each node $v \in \mathcal{V}$ represents a road intersection, a charging station, or the origin of a user. An edge $e\in \mathcal{E}\subseteq \mathcal{V}\times \mathcal{V}$ represents a road connecting two nodes in $\mathcal{V}$. In addition, $\mathcal{N}$ and $\mathcal{L}$ denote the set of buses and the set of distribution lines, respectively. The buses in $\mathcal{N}$ may have a distributed generation unit and load demands (such as charging stations and residential loads). 
Each charging station $d$ from the set of charging stations $\mathcal{D}\subset \mathcal{V}$, which is located at a node $\mathrm{v}$ of the transportation network, imposes an electrical load to the corresponding bus node, say $j\in \mathcal{N}$, of the distribution network.
Fig. \ref{Map_power} demonstrates the proposed system.
\\
The DSO solves a linearized AC optimal power flow (AC-OPF) problem to minimize distributed generators’ operating costs under power system constraints. By solving the linearized AC-OPF, the DLMP at each bus is obtained, which clears the electricity price at charging stations.\\
In the transportation network, there are $M$ different EVs whose set is denoted by $\mathcal{M}$. Each EV $i\in \mathcal{M}$ aims to {compute} the optimum joint routing and destination strategy by considering the traveling time, the price of electricity and service at the charging station, as well as the individual and coupling constraints to reach a charging station located on the transportation and distribution networks.\\
We also consider an aggregator for the transportation network that provides information on traffic conditions to assist EVs in making optimal decisions. 
From the DSO’s point of view, EVs can be considered as mobile electrical loads. The DSO can partially manage such a load by setting charging prices at the stations. 
Therefore, we deal with the problem of joint routing and destination planning of EV users in interdependent power distribution and transportation networks. The high-level information flow of the scheme is depicted in Fig. \ref{Flow}. 
\subsection{Joint Routing and Destination Planning of‌ EVs}
{In this subsection, we introduce each EVs' decision variables, explicit cost function, as well as local constraints.}{
\subsubsection{Decision Variables}
For the EV $i$ with inelastic energy demand ${{q}_{i}}\in \left[ \underline{q},\ \overline{q} \right]$, ${{r}_{i}}=\left[ r_{i}^{e} \right]_{e=1}^{E}$ and ${{t}_{i}}=\left[ t_{i}^{d} \right]_{d=1}^{D}$ are the {decision variables} of choosing roads and charging stations, where $r_{i}^{e}$ and $t_{i}^{d}$ are the probability of selecting road $e$ and charging station $d$, respectively. In addition, ${{r}^{e}}=\left[ r_{i}^{e} \right]_{i=1}^{M}$ and ${{t}^{d}}=\left[ t_{i}^{d} \right]_{i=1}^{M}$ are collective vector of {decision variables} of choosing road $e$ and station $d$, respectively.}
\subsubsection{Cost Function}
The cost function \eqref{cost_EV_opt_prob} is composed of the cost of deviating from preferred route and charging station, the traveling cost, and the cost of station's congestion denoted by $U_i$, $C_i^\mathrm{travel}$, and $C_i^\mathrm{station}$, respectively. {So, we have:
	\begin{equation}\label{cost_EV_opt_prob}
		{J}_{i}={{U}_{i}}+{{\omega }_{i}} C_{i}^\mathrm{travel}+ C_{i}^\mathrm{station},
\end{equation}}
where ${{\omega }_{i}}\in\left[ \underline{\omega },\ \overline{\omega } \right]$ is conversion coefficient of traveling time to a monetary value for user $i$.
The more explicit term for users' preference is:
\begin{equation}
	{{U}_{i}\left( {{r}_{i}},{{t}_{i}} \right)}=\frac{{{\alpha }_{i}}}{2}{{\left\| {{r}_{i}}-{{{\tilde{r}}}_{i}} \right\|}^{2}}+\frac{{{\beta }_{i}}}{2}{{\left\| {{t}_{i}}-{{{\tilde{t}}}_{i}} \right\|}^{2}}.
\end{equation}
Based on user $i$'s previous travel experiences, ${{\tilde{t}}_{i}}={{\left[\tilde{t}_{i}^{d} \right]}_{d=1}^{D}}\in {{\left[ 0,\,1 \right]}^{D}}$, $\sum\limits_{d=1}^{D}{\tilde{t}_{i}^{d}=1}$ and ${{\tilde{r}}^{i}}={{\left[\tilde{r}_{i}^{e} \right]}_{e=1}^{E}}\in {{\left[ 0,\,1 \right]}^{E}}$ represent the preferred destination and path respectively, and ${{\alpha }_{i}}\in[\underline\alpha,\ \bar\alpha]$ and ${{\beta }_{i}}\in[\underline\beta,\ \bar\beta]$ are the constant weight parameters that represent the monetary value of their preferences.	
In addition, we assume that each road segment $e\in {{\mathcal{E}}}$ has a travel time ${{l}_{e}}\left( \cdot  \right)$, which is a continuous and strictly increasing function of the corresponding road's vehicle flow. {A heuristic delay function that is proportionate to the link's traffic level is used to calculate the travel time along each link. The Bureau of Public Roads (BPR) delay model is a well-known latency function that is often used in the literature \cite{Bureau}, which is as follows:}
\begin{equation}\label{travel_time_function}
	{{l}_{e}}({{\sigma }_{e}}({{r}^{e}}))={{\eta}_{e}}\left( 1+\pi {{\left( \frac{{{\sigma }_{e}}({{r}^{e}})}{{{\kappa}_{e}}} \right)}^{\xi }} \right),
\end{equation}
where ${{\eta}_{e}}$ is the free-flow travel time on the road $e$, and ${{\kappa}_{e}}$ is the capacity of road $e$. The parameters $\pi \ge 0$ and $\xi \ge 0$ {are two positive constants of the BPR latency function} \cite{Bureau}. {Parameters $\pi$ and $\xi$ are commonly set to 0.15 and 4, respectively.}
{Also, ${{\sigma }_{e}}\left( {{r}^{e}}  \right)=\sum\limits_{i\in \mathcal{M}}{r_{i}^{e}}+{{s}_{e}}$ is the link flow that is equal to the total expected number of EVs ($\sum\limits_{i\in \mathcal{M}}{r_{i}^{e}}$) and non-EVs (${{s}_{e}}$) on each road segment $e\in {{\mathcal{E}}}$. We define the vector $\sigma \left( r \right)=\left[ {{\sigma }_{e}} \right]_{e=1}^{E}$. In addition, $r=\left[ {{r}_{i}} \right]_{i=1}^{M}$ and $t=\left[ {{t}_{i}} \right]_{i=1}^{M}$ show collective {decision variables} of all EVs regarding choosing the route and destination, respectively.}
Thus, the expected traveling time of the user $i\in \mathcal{M}$ is given by:
\begin{equation}
	C_{i}^\mathrm{travel}\left( {{r}_{i}},\sigma \left( r \right) \right)=\sum\limits_{e\in {{\mathcal{E}}}}{{{l}_{e}}\left( {{\sigma }_{e}}\left( {{r}^{e}} \right) \right)r_{i}^{e}},
\end{equation}\\
We define congestion pricing at the stations to handle overcrowding at the stations considering the growing number of EVs, the limited supply of EV chargers in the stations, as well as the operation cost of the chargers.
	Considering the congestion price ${{\psi }_{d}}$ at station $d$ proportional to the average utilization of the stations, we have:
	\begin{equation}
		{{\psi }_{d}}( {{t}^{d}})=\zeta_d \frac{\delta_d ({{t}^{d}})}{{{\gamma }_{d}}\times {{\phi }_{d}}},
	\end{equation}
	in which for the station $d$, {$\delta_d ({{t}^{d}})=\sum\limits_{i\in \mathcal{M}}{t_{i}^{d}}$} is the arrival rate of EVs, ${{\phi}_{d}}$ is the number of EV chargers, and each charger gets available with the rate ${{\gamma }_{d}}$ (1/(average charging time)), which depends on the charging mode at the station $d$ \cite{reid2019operations}. Finally, $\zeta_d$ is a constant value (in \$) of the operational expenses that is set by charging stations. Also, the collective vector of stations arrival rate is equal to $\delta=\left[ {{\delta }_{d}} \right]_{d=1}^{D}$. {Thus, the expected cost of congestion at the stations for user $i$ is as follows:}
	\begin{equation}
		C_{i}^\mathrm{station}\left( {{t}_{i}},\delta \left( t \right) \right)=\sum\limits_{d\in {{\mathcal{D}}}}{{{\psi}_{d}}\left( {{\delta }_{d}}\left( {{t}^{d}} \right) \right)t_{i}^{d}}.
\end{equation}
\subsubsection{Local Constraint}
Furthermore, for the user $i$, the following linear constraint ensures that the user starting from the initial location ${{o}_{i}}\in \mathcal{V}$ reaches the desired destination $d\in \mathcal{D}$ with probability $t_{i}^{d}$ by considering the underlying transportation graph structure ${{G}_{T}}$.
\begin{equation}\label{Network_cons}
	\begin{split}
		\sum\limits_{e:\left( \mathrm{u},\mathrm{v} \right)\in \mathcal{E}}{r_{i}^{e}}-\sum\limits_{e:\left( \mathrm{v},\mathrm{w} \right)\in \mathcal{E}}{r_{i}^{e}=}
		\begin{cases}
			\ -1 & \text{if} \quad \mathrm{v}={{o}_{i}} \\
			\ t_{i}^{d} & \text{if} \quad \mathrm{v}= d \\
			\ 0 & \text{otherwise}
		\end{cases}\\
		\quad \forall \,\mathrm{v}\in \mathcal{V},\forall \,d\in {\mathcal{D}},
	\end{split}
\end{equation} 
{ where $e:\left( \mathrm{u},\mathrm{v} \right)\in \mathcal{E}$ denotes the link $e\in \mathcal{E}\subseteq \mathcal{V}\times \mathcal{V}$, which originates at node $\mathrm{u}$ and terminates at node $\mathrm{v}$.} As a result, by defining {$x_i = \text{col}(r_i, t_i)$}, each user's individual constraint set can be specified as follows:
\begin{equation}\label{Individual_constraint}
	{\mathcal{X}}_{i}:=\{ {x}_{i}\in {[0,1]}^{E+D}| \text{constraint} \ \eqref{Network_cons},t_{i}^{d}=0\ \forall d\in \mathcal{D}\setminus \mathcal{D}_{i} \},
\end{equation} 
where ${{\mathcal{D}}_{i}}\subseteq \mathcal{D}$ is a set of feasible destinations for EV $i$.
\subsection{OPF problem of the DSO}
{The DSO operates the distribution power network and clears the dispatch level of DGs, real-time electricity prices, active and reactive power flow of lines, and voltages of nodes. In details, $p_h^\mathrm{gen}$, $q_h^\mathrm{gen}$, and $\lambda_\mathrm{dlmp}^j$ show the active and reactive dispatch level of DG $h$, and electricity price at bus $j$, respectively.} {For each line $\left( u,j \right)\in \mathcal{L}$, ${{P}_{uj}}$ and ${{Q}_{uj}}$ are active and reactive power flowing from bus $u$ to bus $j$, respectively. Each line $\left( u,j \right)\in \mathcal{L}$ has its own internal resistance, denoted by $R_{uj}$, reactance, denoted by $X_{uj}$, and power limit, denoted by $S^{\text{max}}_{uj}$. $V_j$ stands for the voltage at node $i$, and we also use $v_j = |V_j|^2$ to compute the linear voltage drop in the system.} In addition, the bus $1$ represents the substation bus, and other buses in $\mathcal{N}$ represent branch buses. {The set of DGs is represented by ${{\Omega }^{G}} = \{1, ..., H\}$.} For each bus $j\in \mathcal{N}$, the set of DGs connected to the bus $j$ is denoted by $\Omega _{j}^{G}\subset {{\Omega }^{G}}$, and $\Omega _{j}^{S}\subset {\mathcal{D}}$ stands for the set of charging stations connected to the bus $j$.
The operational cost of DGs is modeled as a common quadratic function as follows:
\begin{equation}
	{{C}^\mathrm{gen}_{h}}\left( p_{h}^\mathrm{gen} \right)={{a}_{h}}{{\left( p_{h}^\mathrm{gen} \right)}^{2}}+{{b}_{h}}\left( p_{h}^\mathrm{gen} \right)+{{c}_{h}},
\end{equation}
where ${{a}_{h}}$, ${{b}_{h}}$ and ${{c}_{h}}$ are positive cost coefficients. 
 EVs' aggregative variable demand in each station $d$ is obtained as the expected demand of users at the corresponding station:
\begin{equation}
	\varphi_d\left( {{t}^{d}} \right)=\sum\limits_{i\in \mathcal{M}}{{{q}_{i}}t_{i}^{d}}.
\end{equation}
Therefore, power demand at bus $j$ is determined by the summation of fixed residential load connected to the bus denoted by $p_{j}^\mathrm{load}$ and variable total expected demand {of EVs} at stations ${d\in {{\Omega }_{j}^{S}}}$ connected to the bus $j$. 
 The DSO uses the \textit{LinDistFlow} model \cite{baran1989optimal}, which is a linear approximation of the AC power flow model, to minimize total generation costs while taking operational constraints into account.
Hence, the optimization problem of the DSO can be expressed as follows:
{\footnotesize
\begin{subequations}
	\label{opf}
	\begin{align}
		&\underset{y}{\mathop{\text{min}}}\,\quad {{J}_{0}}\left( {{p}^\mathrm{gen}} \right)=\sum\limits_{h\in {{\Omega }^{G}}}{{{C}^\mathrm{gen}_{h}}\left( p_{h}^\mathrm{gen} \right)}\label{cost__gen}\\
		\begin{split}
		&\text{s.t.}\sum\limits_{u:\left( j,u \right)\in \mathcal{L}}{{{P}_{ju}}} - {{{P}_{uj}}}=\sum\limits_{h\in \Omega _{j}^{G}}p_{h}^\mathrm{gen}-p_{j}^\mathrm{load}-\sum\limits_{d\in \Omega _{j}^{S}}{\varphi_d( {{t}^{d}} )},\forall j\in \mathcal{N}
		\end{split}\label{power_balance}\\
		&\quad\sum\limits_{u:\left( j,u \right)\in \mathcal{L}}{{{Q}_{ju}}} - {{{Q}_{uj}}}=\sum\limits_{h\in \Omega _{j}^{G}}q_{h}^\mathrm{gen}-q_{j}^\mathrm{load}\quad,\forall j\in \mathcal{N}\label{power_balance_q}\\
		&\quad \ v_j = v_u - 2R_{uj}P_{uj} - 2X_{uj}Q_{uj}\quad\quad\quad\quad\quad\ ,\forall \left( u,j \right)\in \mathcal{L}\label{voltage_drop}\\
		&\quad \ \underline{p}_{h}^\mathrm{gen}\le p_{h}^\mathrm{gen}\le \overline{p}_{h}^\mathrm{gen}\quad\quad\quad\quad\quad\quad\quad\quad\quad\quad\quad  ,\forall h\in {{\Omega }^{G}}\label{generation_constraint}\\
		&\quad \ \underline{q}_{h}^\mathrm{gen}\le q_{h}^\mathrm{gen}\le \overline{q}_{h}^\mathrm{gen}\quad\quad\quad\quad\quad\quad\quad\quad\quad\quad\quad,\forall h\in {{\Omega }^{G}}\label{generation_constraint_q}\\
		&\quad \ \underline{v}_{j}\le v_{j}\le \overline{v}_{j}\quad\quad\quad\quad\quad\quad\quad\quad\quad\quad\quad\quad\quad\ \ ,\forall j\in \mathcal{N}\label{voltage_constraint}\\
		&\quad \ {{P}^2_{uj}} + {{Q}^2_{uj}}\le {{{S}}^{\text{max}}_{uj}}\quad\quad\quad\quad\quad\quad\quad\quad\quad\quad\quad\ ,\forall \left( u,j \right)\in \mathcal{L}. \label{line_constraint}
	\end{align}
\end{subequations}}
{We let $y=\text{col}\left( {p}^\mathrm{gen}, {q}^\mathrm{gen}, {{P}_{L}}, {{Q}_{L}}, v \right)$ stand for the DSO's collective decision variable, in which ${{p}^\mathrm{gen}}=\left[ p_{h}^\mathrm{gen} \right]_{h=1}^{H}$, ${{q}^\mathrm{gen}}=\left[ q_{h}^\mathrm{gen} \right]_{h=1}^{H}$, ${{v}}=\left[ p_{j} \right]_{j=1}^{\left| \mathcal{L} \right|}$, ${{P}_{L}}=\left\{ {{P}_{uj}}|\ \forall\left( u,j \right)\in \mathcal{L} \right\}$, ${{Q}_{L}}=\left\{ {{Q}_{uj}}|\ \forall\left( u,j \right)\in \mathcal{L} \right\}$. $\overline{p}_{h}^\mathrm{gen}$ and $\underline{p}_{h}^\mathrm{gen}$ are the maximum and minimum active power outputs of DG $h$. Similarly, $\overline{q}_{h}^\mathrm{gen}$ and $\underline{q}_{h}^\mathrm{gen}$ are the maximum and minimum reactive power outputs of DG $h$. Lastly, $\overline{v}_{j}$ and $\underline{v}_{j}$ are the maximum and minimum values for voltage constraint of node $j$.}
{Equation \eqref{power_balance} enforces active power balance at the distribution network, whereas equation \eqref{power_balance_q} enforces reactive power balance. Equation \eqref{voltage_drop} provides a linear voltage drop relationship across the distribution system. Moreover, \eqref{generation_constraint} and \eqref{generation_constraint_q} represent DGs' active and reactive power output constraints. Lastly, \eqref{voltage_constraint} are nodal voltage constraints and \eqref{line_constraint} corresponds to the power constraints for each distribution line.}\\
{The equality constraint \eqref{power_balance} is best described as a coupling constraint between users and the DSO, as the decisions of all users affect the power balance equation of the optimal power flow (OPF) problem. The feasible set corresponding to the constraint \eqref{power_balance} can be written as follows:
	\begin{equation}\label{cons_cons}
		{{\mathcal{C}}^\mathrm{power}}=\left\{ t\in {{\mathbb{R}}^{MD}},y\in {{\mathbb{R}}^{2H+3\left| \mathcal{L} \right|}}\left|\  \text{constraint}\ \eqref{power_balance} \right. \right\},
	\end{equation}
	where ${{\lambda }_\mathrm{dlmp}}=[\lambda ^{j}_\mathrm{dlmp}]_{j=1}^{N}$ is the Lagrangian multiplier vector of the constraint \eqref{cons_cons}. In the power balance equation \eqref{power_balance}, the effect of EVs' variable expected demand is shown. Also, for a station $d$ that is connected to the bus $j$, the per-unit electricity price ${\lambda}^{d}_\mathrm{station}$ for the station is equal to the DLMP price $\lambda ^{j}_\mathrm{dlmp}$ for the bus $j$, which is derived from solving problem \eqref{opf}; so we have:
	\begin{equation}
		{\lambda}^{d}_\mathrm{station}=\lambda ^{j}_\mathrm{dlmp},\quad \forall d\in \Omega _{j}^{S}.
	\end{equation}
If the distribution network is not congested, DLMPs in {all} buses and, subsequently, prices at {all} the charging stations will be the same. As the demand for certain buses increases and the power network's constraints get violated, prices at those buses increase in real-time to encourage users to switch to charging stations with un-congested buses or cheaper generation units.

	The DSO's private constraint set is the feasible set that corresponds to the constraints \eqref{power_balance_q}-\eqref{line_constraint}, which is:
	\begin{equation}
		\mathcal{Y}=\left\{ y\in {{\mathbb{R}}^{2H+3\left| \mathcal{L} \right|}}\left| \ \text{constraints}\ \eqref{power_balance_q}-\eqref{line_constraint}\right. \right\}.
\end{equation}}
The resulted convex optimization problem \eqref{opf} can be solved in real-time with acceptable accuracy \cite{zhu2015fast}.
\subsection{Coupling Constraints}
{ We consider the coupling constraints among users related to stations' power capacity and transportation network vehicles' flow limit as follows:
	\begin{subequations}
		\begin{align}
			{{\mathcal{C}}^{t}}=\left\{ t\in {{\mathbb{R}}^{MD}}\left| \varphi_d( {{t}^{d}})\le {{{c}}^t_{d}},\forall d\in \mathcal{D} \right. \right\}\label{station_cap_cons}\\
			{{\mathcal{C}}^{r}}=\left\{ r\in {{\mathbb{R}}^{ME}}\left| {{\sigma }_{e}}({{r}^{e}})\le {{{c}}^r_{e}},\forall e\in \mathcal{E} \right. \right\}\label{road_cap_con}.		
		\end{align}
	\end{subequations}
	The constraint \eqref{station_cap_cons} with Lagrangian multipliers ${{\lambda }_{t}}=[\lambda ^{d}_{t}]_{d=1}^{D}$ ensures that, for each station $d$, the expected charging demand must be less than the station capacity's upper bound ${{{c}}^t_{d}}$.
	Additionally, constraint \eqref{road_cap_con} with Lagrangian multipliers ${{\lambda }_{r}}=[\lambda ^{e}_{r}]_{e=1}^{E}$ indicates that each road segment $e$ cannot serve more vehicles than its upper bound ${{{c}}^r_{e}}$. Stations' capacity vectors and roads' capacity vectors are denoted by ${{c}^{t}}=\left[ {c}_{d}^{t} \right]_{d=1}^{D}$ and ${{c}^{r}=\left[ {c}_{e}^{r} \right]_{e=1}^{E}}$, respectively.\\
	Also, on the basis of what was mentioned in the previous two subsections, we define the collective coupling constraint on the system as follows:
	\begin{equation}\label{coup_constraint}
		\mathcal{C}={{\mathcal{C}}^{r}}\times ( {{\mathcal{C}}^{t}}\bigcap {{\mathcal{C}}^\mathrm{power}} )
\end{equation}}
{
	In fact, Lagrange multipliers ${{\lambda }_\mathrm{dlmp}}$, ${{\lambda }_{t}}$, and ${{\lambda }_{r}}$ are prices imposed by the DSO, CSOs, and the TNO, respectively, to influence consumers.}
\section{Decentralized Game-Theoretical Analysis}\label{Decentralized}
Given the users' competition for the fastest route to the least crowded stations and the limits on charging stations, roads, and power system, the interaction among EV users and DSO can be formally defined as a game problem.
\subsection{Game formulation}
Based on the introduced models, an aggregative game with coupling constraints can be defined among EV users and the DSO -the game $\mathcal{G}$- whose elements are as follows:

{
\begin{equation}\label{Game_power}
	\mathcal{G}=\left\{ \begin{aligned}
		& \mathbf{Players}\ :M\ \text{EVs and the DSO}\\ 
		& \mathbf{Strategies}\ :\left\{ \begin{aligned}
			& \text{EV}\ i:{{x}_{i}}\in {{\mathcal{X}}_{i}} \\ 
			& \text{DSO}\ :\ y\in \mathcal{Y} \\ 
		\end{aligned} \right. \\ 
		& \mathbf{Costs}\ :\left\{ \begin{aligned}
			& \text{EV}\ i:{{J}_{i}}\left( {{x}_{i}},\Theta \left( x \right) \right) \\ 
			& \text{DSO}:\ {{J}_{0}}\left( y \right) \\ 
		\end{aligned} \right. \\ 
		& \mathbf{Coupling}\ \mathbf{Constraints}\ :\mathcal{C} \\ 
	\end{aligned} \right. .
\end{equation}}

We also let $\mathcal{K}:=( \prod\limits_{i=1}^{M}{{{\mathcal{X}}_{i}}\times \mathcal{Y}} )\bigcap \mathcal{C}$ denote the cumulative strategy set of the game $\mathcal{G}$.\\
The concept of Wardrop equilibrium is well known in transportation networks \cite{wardrop1952road}. Specifically, we consider the variational generalized Wardrop equilibrium \cite{14,16}.
\begin{definition}(\textit{generalized Wardrop equilibrium})
	A collective strategy $\text{col}\left( \bar{x},\bar{y} \right)$ is called a GWE of the game $\mathcal{G}$ if for any $i \in \mathcal{M}$ we have:
	{
	\begin{equation}
	\begin{split}
		&{{J}_{i}}\left( {{{\bar{x}}}_{i}},\Theta \left( {\bar{x}} \right) \right)\le {{J}_{i}}\left( {{x}_{i}},\Theta \left( {{{\bar{x}}}} \right) \right), \\ 
		&\forall x_i\ \text{s.t.}\ {{x}_{i}} \in \mathcal{K}\left( {{{\bar{x}}}_{-i}},\bar{y} \right), \\
		&{{J}_{0}}\left( {\bar{y}} \right)\le {{J}_{0}}\left( y \right), \forall y\ \text{s.t.}\ y \in \mathcal{K}\left( \bar{x} \right), 
	\end{split}
	\end{equation}}
	where $x=\left[ {{x}_{i}} \right]_{i=1}^{M}$, $\bar{x}=\left[{{\bar{x}}_{i}} \right]_{i=1}^{M}$, $\Theta(x) = \text{col} (\sigma (r), \delta(t))$, and ${{\bar{x}}_{-i}}=\left( {{{\bar{x}}}_{1}},\ldots ,{{{\bar{x}}}_{i-1}},{{{\bar{x}}}_{i+1}},\ldots ,{{{\bar{x}}}_{M}} \right).$\\
	In other words, GWE is a set of decisions in which no agent can reduce its cost function unilaterally by modifying its decision within the feasible constraint set, given that the aggregated {strategies $\Theta(x)$ are} fixed.
\end{definition}
\begin{definition}
 { The pseudo-gradient mapping of game $\mathcal{G}$ is defined by stacking together the gradient of players' cost functions with respect to their local decision variables, while $\Theta \left( x \right)$ is fixed, as follows:
\begin{equation}\label{game_mapping}
	F\left( x,y \right):=\left[ \begin{aligned}
		& \left[ {{\nabla }_{{{x}_{i}}}}{{J}_{i}}\left( {{x}_{i}},z \right)\left| _{z=\Theta (x)} \right. \right]_{i=1}^{M} \\ 
		& \quad\quad\quad{{\nabla }_{y}}{{J}_{0}}\left( y \right) \\ 
	\end{aligned} \right].
\end{equation}}
\end{definition}
In order to ensure the satisfaction of coupling constraint {\eqref{coup_constraint}}, we postulate that the Lagrange multipliers {$\lambda =\text{col}\left( {{\lambda }_\mathrm{dlmp}},{{\lambda }_{r}},{{\lambda }_{t}} \right)$} are the same for all EVs. By this assumption, we restrict the analysis to a subset of equilibrium points of $\mathcal{G}$, which is known as the variational generalized Wardrop equilibrium (v-GWE) {\cite{16}}.
{
\begin{definition}(\textit{Variational Inequality})
	Given a set $\mathcal{K}$ and mapping $F:\mathcal{K}\to {{\mathbb{R}}^{n}}$,
	the variational inequality problem $\text{VI}\left( F,\mathcal{K} \right)$, is to find a vector $\text{col}\left( \bar{x},\bar{y} \right)\in \mathcal{K}$ such that:
	\begin{equation}
		\label{vari}
		F{{\left( \bar{x},\bar{y} \right)}^{\top }}\left(\text{col}\left( {x},{y} \right) -\text{col}\left( \bar{x},\bar{y} \right)\right)\ge 0,\ \forall \ \text{col}\left( {x},{y} \right)\in \mathcal{K}
	\end{equation}
The set of solutions to \eqref{vari} is denoted by $\text{SOL}\left( F,\mathcal{K} \right)$.
\end{definition}}
In this case, the game $\mathcal{G}$ can be characterized as a variational inequality problem $\text{VI}\left( F,\mathcal{K} \right)$. This representation helps to obtain the v-GWEs of $\mathcal{G}$ by solving the corresponding variational inequality problem. 
The relation between $\text{SOL}\left( F,\mathcal{K} \right)$ and the v-GWEs of $\mathcal{G}$ is established in the next Proposition.
\begin{proposition}\label{solution_game}
For game $\mathcal{G}$, $\text{SOL}\left( F,\mathcal{K} \right)$ of $\text{VI}\left( F,\mathcal{K} \right)$ is also a v-GWE of game $\mathcal{G}$.
\end{proposition}
\begin{IEEEproof}
	For each user $i \in \mathcal{M}$ and the DSO, the set $\mathcal{X}_i$ and $\mathcal{Y}$ are non-empty, closed, and convex. Also for each $i$, function ${{J}_{i}}({{x}_{i}},\Theta(x))$ is continuously differentiable and convex in $x_i$, and function $J_0(y)$ is continuously differentiable and convex in $y$. The set $\mathcal{C}$ is also non-empty and satisfies Slater’s constraint qualification. Therefore, due to \cite[Proposition 12.4]{Palomar2009}, any solution of $\text{SOL}\left( F,\mathcal{K} \right)$ is a v-GWE of $\mathcal{G}$.
\end{IEEEproof}
\begin{definition}
Mapping $F:\mathcal{X}\in {{\mathbb{R}}^{n}}\to {{\mathbb{R}}^{n}}$ is monotone on $\mathcal{X}$ if, for every $x,y\in \mathcal{X},x\ne y$, ${{\left( F( x )-F( y ) \right)}^{\top }}\left( x-y \right)\ge 0 $.
\end{definition}
\begin{lemma}\label{monotinicity}
	Operator $F(x,y)$ in \eqref{game_mapping} is monotone.
\end{lemma}
\begin{IEEEproof}
	We can show operator $F(x,y)$ as follows:
	\begin{equation}
		\begin{split}
			&F={\left[ \begin{aligned}
					& \left[ \begin{aligned}
						& {{\alpha }_{i}}({{r}_{i}}-{{{\tilde{r}}}_{i}})+{{\omega }_{i}}\ l(\sigma (r)) \\ 
						& {{\beta }_{i}}({{t}_{i}}-{{{\tilde{t}}}_{i}})+\ \psi(\delta (t)) \\ 
					\end{aligned} \right]_{i=1}^{M} \\ 
					& \left[ \begin{aligned}
						& \left[ 2{{a}_{h}}\left( p_{h}^\mathrm{gen} \right)+{{b}_{h}} \right]_{h=1}^{H} \\ 
						& {{0}_{\left| \mathcal{L} \right|\times 1}} \\ 
					\end{aligned} \right] \\ 
				\end{aligned} \right]}
		\end{split},
	\end{equation}
	where $l(\sigma (r))=\left[ {{l}_{e}}({{\sigma }_{e}}) \right]_{e=1}^{E}$ and $\psi(\delta (t))=\left[ {{\psi }_{d}}({{\delta }_{d}}) \right]_{d=1}^{D}$.
	Since functions $U_i(x_i)$, $l(\sigma (r))$, and $\psi(\delta (t))$ are convex in $x$ and feasible sets for each $i \in \mathcal{M}$, and function $J_0(y)$ is convex in $y$, the first term is monotone. Consequently, the mapping $F\left( x,y \right)$ is monotone.
\end{IEEEproof}
\begin{theorem}\label{existence}
	The solution set $\text{SOL}\left( F,\mathcal{K} \right)$ is non-empty and compact; therefore, v-GWE exists.
\end{theorem}
\begin{IEEEproof}
	{Since for all $i \in \mathcal{M}$ the sets $\mathcal{X}_i$ and $\mathcal{Y}$ are closed, convex, and bounded, the function ${{J}_{i}}({{x}_{i}},\Theta(x))$ is continuously differentiable and convex in $x_i$, the function $J_0(y)$ is continuously differentiable and convex in $y$ and the set $\mathcal{C}$ is non-empty; therefore, \cite[Corollary 2.2.5]{114} guarantees the existence of an equilibrium of game $\mathcal{G}$.}
\end{IEEEproof}
\subsection{Decentralized Algorithm}
{In this subsection, we want to solve the game \eqref{Game_power} in a decentralized way. The optimization subproblem of each EV can be derived by relaxing coupling constraints $\mathcal{C}$ as follows:

{\footnotesize
\begin{equation}\label{cost_aug}
	\underset{{{x}_{i}}\in {{\mathcal{X}}_{i}}}{\mathop{\min }}\,\ {{\tilde{J}}_{i}}({{r}_{i}},{{t}_{i}},\Theta (x),\lambda )=\underbrace{{{J}_{i}}}_{\text{term 1}}+\underbrace{\sum\limits_{e=1}^{E}{\lambda _{r}^{e}}r_{i}^{e}}_{\text{term 2}}+\underbrace{\sum\limits_{d=1}^{D}{{{q}_{i}}(\lambda _{t}^{d}+\lambda _{\text{station}}^{d})t_{i}^{d}}}_{\text{term 3}},
\end{equation}}

{where $x=\left[ {{x}_{i}} \right]_{i=1}^{M}$, $\Theta(x) = \text{col} (\sigma (r), \delta(t))$, and $\lambda =\text{col}\left( {{\lambda }_\mathrm{dlmp}},{{\lambda }_{r}},{{\lambda }_{t}} \right)$. {In summary, the first term in the cost function \eqref{cost_aug} reflects the explicit cost function of EVs mentioned in \eqref{cost_EV_opt_prob}. The second term denotes the user's payment of tolls for road usage, the third term shows the cost of energy used to charge the EV.}}
To solve the problem in a decentralized manner, we introduce the following extended single-valued mapping $T$ with variables $z=\text{col}\left( x,y,\lambda  \right)$:}

{\footnotesize
	\begin{subequations}
	\begin{align}\label{T_1}
		&T=\left[ \begin{aligned}
			&F+\left[ \begin{aligned}
				& \left[ {{\nabla }_{{{x}_{i}}}}( {{q}_{i}}( {{t}_{i}}^{\top }{{\lambda }_{t}}+{{t}_{i}}^{\top }{{\lambda }_\mathrm{station}} )+{{r}_{i}}^{\top }{{\lambda }_{r}} ) \right]_{i=1}^{M} \\ 
				& {{\nabla }_{y}}( \sum\limits_{j\in \mathcal{M}}{\lambda ^{j}_\mathrm{dlmp}(\sum\limits_{\left( j,u \right)\in \mathcal{L}}{{{P}_{ju}}}-{{{P}_{uj}}}-\sum\limits_{h\in \Omega _{j}^\mathrm{gen}}{p_{h}^\mathrm{gen}})} ) \\ 
			\end{aligned} \right]  \\ 
			&\quad\quad\quad\quad\quad\quad\quad \left[ -EC\left( t,y \right)-{{p}^\mathrm{load}} \right] \\ 
			&\quad\quad\quad\quad\quad\quad\quad\quad \left[ {{{c}}^r_{d}}-\varphi_d ( {{t}^{d}} ) \right]_{d=1}^{D} \\ 
			&\quad\quad\quad\quad\quad\quad\quad\quad \left[ {{{c}}^t_{e}}-{{\sigma }_{e}}\left( {{r}^{e}} \right) \right]_{e=1}^{E} \\ 
		\end{aligned} \right]\\
	&\quad\quad\quad\quad\quad\mathcal{Z}:=\prod\limits_{i=1}^{M}{{{\mathcal{X}}_{i}}\times \mathcal{Y}\times {{\mathbb{R}}^{N}}\times \mathbb{R}_{\ge 0}^{E}\times \mathbb{R}_{\ge 0}^{D}},
	\end{align}
\end{subequations}}

where $EC\left( t,y \right)$ is the corresponding expression in the constraint \eqref{power_balance} as follows:

{\footnotesize
\begin{equation}
	EC\left( t,y \right)=\left[ \sum\limits_{\left( j,u \right)\in \mathcal{L}}{{{P}_{ju}}} - {{{P}_{uj}}}-\sum\limits_{h\in \Omega _{j}^\mathrm{gen}}{p_{h}^\mathrm{gen}+\sum\limits_{d\in \Omega _{j}^{s}}{\varphi_d ( {{t}^{d}} )}} \right]_{j=1}^{N}.
\end{equation}}

The zeros of mapping $T$ are equal to the equilibrium of the game $\mathcal{G}$ in \eqref{Game_power}. Since the mapping $T$ is decomposable, it can be utilized to seek the v-GWE in a decentralized way. {It is worth noting that the pseudo gradients associated with EVs in equation \eqref{T_1} are gradients of functions in equation \eqref{cost_aug}.} 
\begin{proposition}
	Two variational inequality problems, $\text{VI}(T,\mathcal{Z})$ and $\text{VI}(F,\mathcal{K})$, are equivalent.
\end{proposition}
\begin{IEEEproof}
	Due to the conditions hold in proposition \ref{solution_game}, and according to \cite[Theorem 3.1]{auslender2000lagrangian}, considering the same Lagrangian multipliers correspond to the coupling constraints $\mathcal{C}$ for the users, $\overline{\lambda}\in \mathbb{R}^{N+E+D}$ exists such that $\text{col}(\overline{x}, \overline\lambda)=\text{SOL}(T,\mathcal{Z})$ if and only if $\overline{x}=\text{SOL}(F,\mathcal{K})$ which is a variational equilibrium of the game $\mathcal{G}$.
\end{IEEEproof}
{To find the equilibrium point $\text{VI}(T,\mathcal{Z})$, we use the I-FoRB splitting method \cite{14} and \cite{15}. The suggested algorithm is shown in algorithm \ref{algorithm}. Unlike most decentralized methods, which require strong monotonicity conditions, this algorithm only requires the monotonicity of mapping $F$ (which was proved in Lemma 1) to ensure the convergence to {a} Wardrop equilibrium point of the game. Furthermore, the majority of the algorithms in the literature use a two-time scale scheme with a vanishing step size, which significantly reduces the convergence rate. {Nevertheless, I-FoRB overcomes this problem by splitting two maximally monotone operators, one of which is Lipschitz continuous and single valued.}
Note that in Algorithm 1, $f_i(x_i,\Theta(x))={{\omega}_{i}}C_{i}^\mathrm{travel}+C_{i}^\mathrm{station}$ is a part of the cost function of EV that is affected by aggregate decision of all users.}
\begin{algorithm} 
	\caption{Aggregative game}
	\begin{algorithmic}	
		\STATE\textbf{Initialization:}
		$k \leftarrow 1$, $\tau_i >0$, $\mu >0$, $\vartheta\in[0,1/3)$, $ t_{i}^{\left( 0 \right), \left( 1 \right)}\in \mathbb{R}_{\ge 0}^{D}$, $ r_{i}^{\left( 0 \right), \left( 1 \right)}\in \mathbb{R}_{\ge 0}^{E}$, $ y^{\left( 0 \right), \left( 1 \right)}\in \mathbb{R}^{2H+3\left| \mathcal{L} \right|}$, $ \lambda^{\left( 0 \right), \left( 1 \right)}_\mathrm{dlmp}\in \mathbb{R}^{N}$, $ \lambda^{\left( 0 \right), \left( 1 \right)}_{t}\in \mathbb{R}_{\ge 0}^{D}$, $ \lambda^{\left( 0 \right), \left( 1 \right)}_{r}\in \mathbb{R}_{\ge 0}^{E}$.
		\STATE\textbf{Iteration} $k$: \vspace{0.1cm}
		\STATE\vline\vspace{0.00cm}
		\begin{minipage}{\linewidth}
			\STATE\textbf{EVs}: $\forall i\in {\cal M}$:
				\[\begin{aligned}
				& GR_{i}^{\left( k \right)}=2{{\nabla }_{{{x}_{i}}}}{{f}_{i}}({{x}_{i}}^{\left( k \right)},z)\left| _{z=\Theta ({{x}^{(k)}})} \right.\\
				&\quad\quad\quad-{{\nabla }_{{{x}_{i}}}}{{f}_{i}}({{x}_{i}}^{\left( k-1 \right)},z)\left| _{z=\Theta ({{x}^{(k-1)}})} \right. \\ 
				& \quad\quad\quad+{{\nabla }_{{{x}_{i}}}}( {{q}_{i}}{{( \lambda _{t}^{(k)}+\lambda _\mathrm{station}^{(k)} )}^{\top }}t_{i}^{\left( k \right)}+{{ \lambda _{r}^{(k)}}^{\top }}r_{i}^{\left( k \right)} )\\
				& {x}_{i}^{\left( k+1 \right)}={{\Pi}_{{{\tau }_{i}}{{U}_{i}}+{{\iota }_{{{\mathcal{X}}_{i}}}}}}[ {x}_{i}^{\left( k \right)}-{{\tau }_{i}}\ GR_{_{i}}^{\left( k \right)}+\vartheta ( x_{i}^{\left( k \right)}-x_{i}^{\left( k-1 \right)})]\\
			\end{aligned}\]
			\STATE\textbf{Power Distribution Network Operator}:
				\[\begin{aligned}
				& GR_{_{0}}^{\left( k \right)}={{\nabla }_{y}}( \lambda _\mathrm{dlmp}^{\top }EC( {{t}^{\left( k \right)}},{{y}^{\left( k \right)}} ) ) \\ 
				& {{y}^{\left( k+1 \right)}}={{\Pi}_{{{\tau }_{0}}{{J}_{0}}+{{\iota }_{\mathcal{Y}}}}}[ {{y}^{\left( k \right)}}-{{\tau }_{0}}\ GR_{_{0}}^{\left( k \right)} +\vartheta ( {{y}^{\left( k \right)}}-{{y}^{\left( k-1 \right)}} )] \\ 
				& R_\mathrm{dlmp}^{\left( k+1 \right)}=2EC( {{y}^{\left( k+1 \right)}},{{t}^{\left( k+1 \right)}} )-EC( {{y}^{\left( k \right)}},{{t}^{\left( k \right)}} )-{{p}^\mathrm{load}} \\ 
				& \lambda _\mathrm{dlmp}^{(k+1)}={\lambda}_\mathrm{dlmp}^{(k)}+\mu R_\mathrm{dlmp}^{\left( k+1 \right)}+\vartheta ( \lambda _\mathrm{dlmp}^{(k)}-\lambda _\mathrm{dlmp}^{(k-1)} ) \\
				&({\lambda ^{d}_\mathrm{station}})^{(k+1)}=({\lambda ^{j}_\mathrm{dlmp}})^{(k+1)},\quad \forall d\in \Omega _{j}^{S}\\\end{aligned}\]
			\STATE\textbf{Transportation Network Operator}: $\forall e \in \mathcal{E}$:
				\[\begin{aligned}
				& {R^r_{e}}^{\left( k+1 \right)}=2\sigma_e ( {{r^e}^{\left( k+1 \right)}} )-\sigma_e ( {{r^e}^{\left( k \right)}} )-{c}_e^r ,\\ 
				&{\lambda^e_{r}}^{(k+1)}={{\text{proj} }_{\mathbb{R}_{\ge 0}}}[ {\lambda^e_{r}}^{(k)}+\mu {R^r_{e}}^{\left( k+1 \right)}+ \vartheta ( {\lambda^{e}_{r}}^{(k)}-{\lambda^e_{r}}^{(k-1)} )] \\
			\end{aligned}\]
			\STATE\textbf{Charging Station Operators}: $\forall d \in \mathcal{D}$:
			\[\begin{aligned}
				& {R^t_{d}}^{\left( k+1 \right)}=2\varphi_d ( {{t^d}^{\left( k+1 \right)}} )-\varphi_d ( {{t^d}^{\left( k \right)}} )-{c}^t_d\\ 
				& {\lambda^d_{t}}^{(k+1)}={{\text{proj} }_{\mathbb{R}_{\ge 0}}}[ {\lambda^d_{t}}^{(k)}+\mu {R^t_{d}}^{\left( k+1 \right)}+\vartheta ( {\lambda^d_{t}}^{(k)}-{\lambda^d_{t}}^{(k-1)} )]\\
			\end{aligned}\]
			\STATE 	$ k \leftarrow k+1$
		\end{minipage}
	\end{algorithmic}
	\label{algorithm}
\end{algorithm}
{The learning step size $\tau_{i}$ is set independently and locally by each entity in the Algorithm \ref{algorithm}. Each EV updates its strategy using the gradient of its cost function based on the most recent common information provided by the aggregators and then projects its strategy onto its feasible local set using the proximal operator. Then, the DSO collects electrical demand data at distribution buses and updates the optimal power flow variables, including DLMPs, by projecting onto its feasible constraint, considering the deviation of load balance constraint and its operational costs, and updating the dual variables. Following that, the prices are broadcast to CSOs and EVs. Meanwhile, each CSO, along with the TNO collects the associated aggregative strategies, adjusts the tolls based on their capacity constraints, and broadcasts the shared data to the EVs. This process is repeated until an equilibrium point is reached.}
In the following proposition, we prove the convergence of the algorithm \ref{algorithm} to a v-GWE of the game.
\begin{proposition}
	Algorithm \ref{algorithm} converges to a solution of $\text{VI}(F,\mathcal{K})$.
\end{proposition}
\begin{IEEEproof}
	As we showed before, due to conditions held in lemma \ref{monotinicity}, proposition \ref{solution_game}, and theorem \ref{existence}, according to \cite[Algorithm 5]{14} if the positive step-sizes $\tau_{i}$, $\tau_{0}$, and $\mu$ are selected sufficiently small, and inertial parameter $\vartheta$ selected within $\vartheta\in [0,1/3)$ for each $i\in\mathcal{M}$, the algorithm \ref{algorithm} would globally converge to $\text{col}(x, \lambda)\in\text{SOL}(T,\mathcal{Z})$ \cite[Theorem 1]{14}.
\end{IEEEproof}
\section{Simulation Results}\label{Simulation}
Our simulations are conducted on Savannah city model that includes 58 bidirectional roads, 36 nodes, and six charging stations. {{We utilize the linear form of the delay function with $\xi=1$ and $\pi=4$ \cite{115}.
The average vehicle speed on a regular day in Savannah is 30 km/h, from which the free flow travel time on roads ($\frac{\text{length}}{\text{free-flow\  speed}}$) is computed. We assume 20 vehicles per kilometer traveling in an uncongested situation as the capacity of roads is heuristically determined by ${{\kappa}_{e}}=30\frac{\text{vehicle}}{\text{km}}\times (\text{free-flow speed})$.} In addition, IEEE 33-bus radial distribution network is considered, which is connected to charging stations positioned on the transportation network.

Figures \ref{Map} and \ref{33_Bus} illustrate the transportation and power networks, respectively. Fig. \ref{33_Bus} also illustrates the locations of distributed generation units and charging stations in the distribution network, where one substation unit and four distributed generation units are assumed.
\begin{figure}[ht]
	\centering
	\includegraphics[width=.7\linewidth]{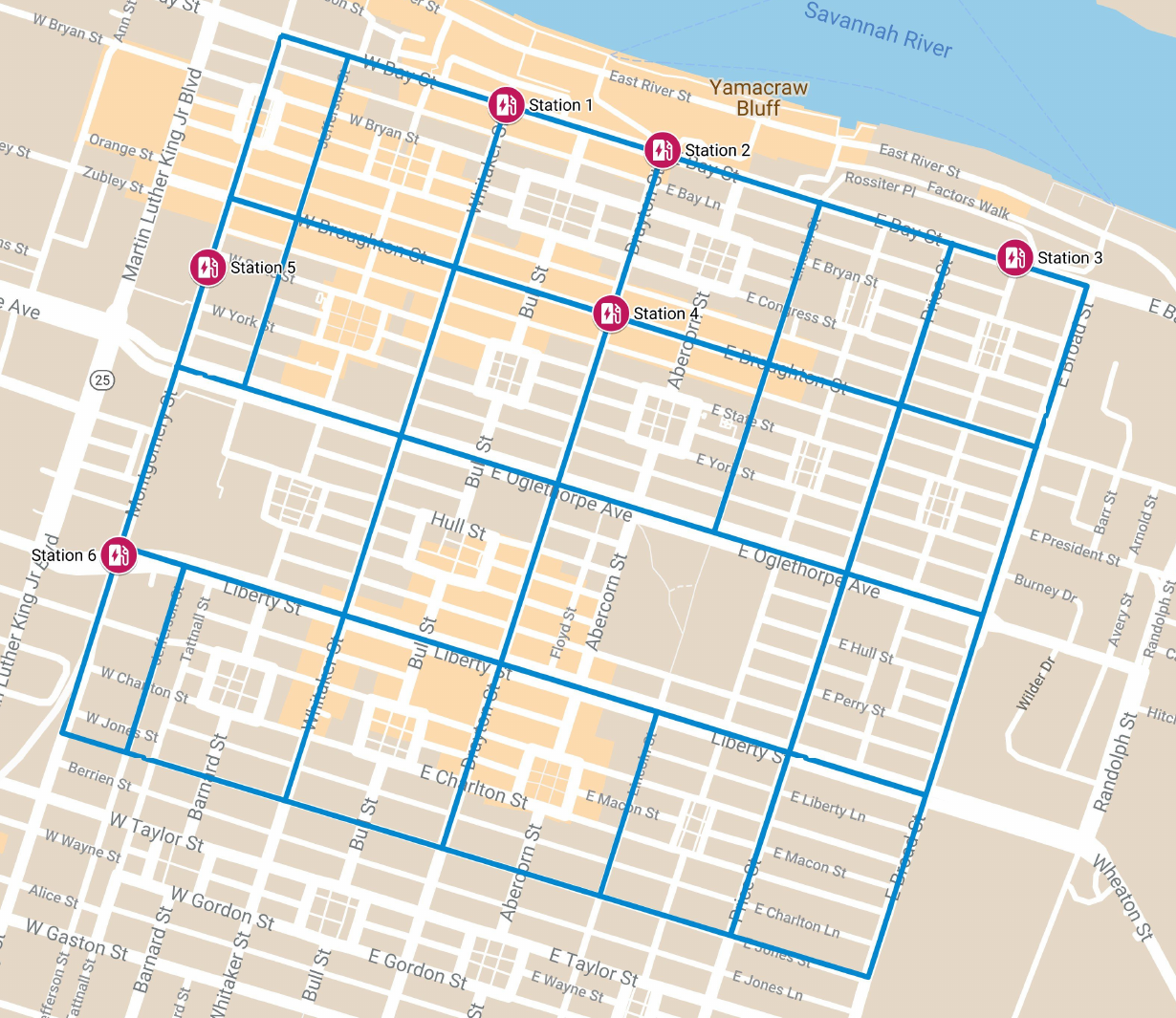}
	\caption{Transportation network of Savannah city, where CSs are located}
	\label{Map}
\end{figure}
\begin{figure}[!ht]
	\centering
	\includegraphics[width=1\linewidth]{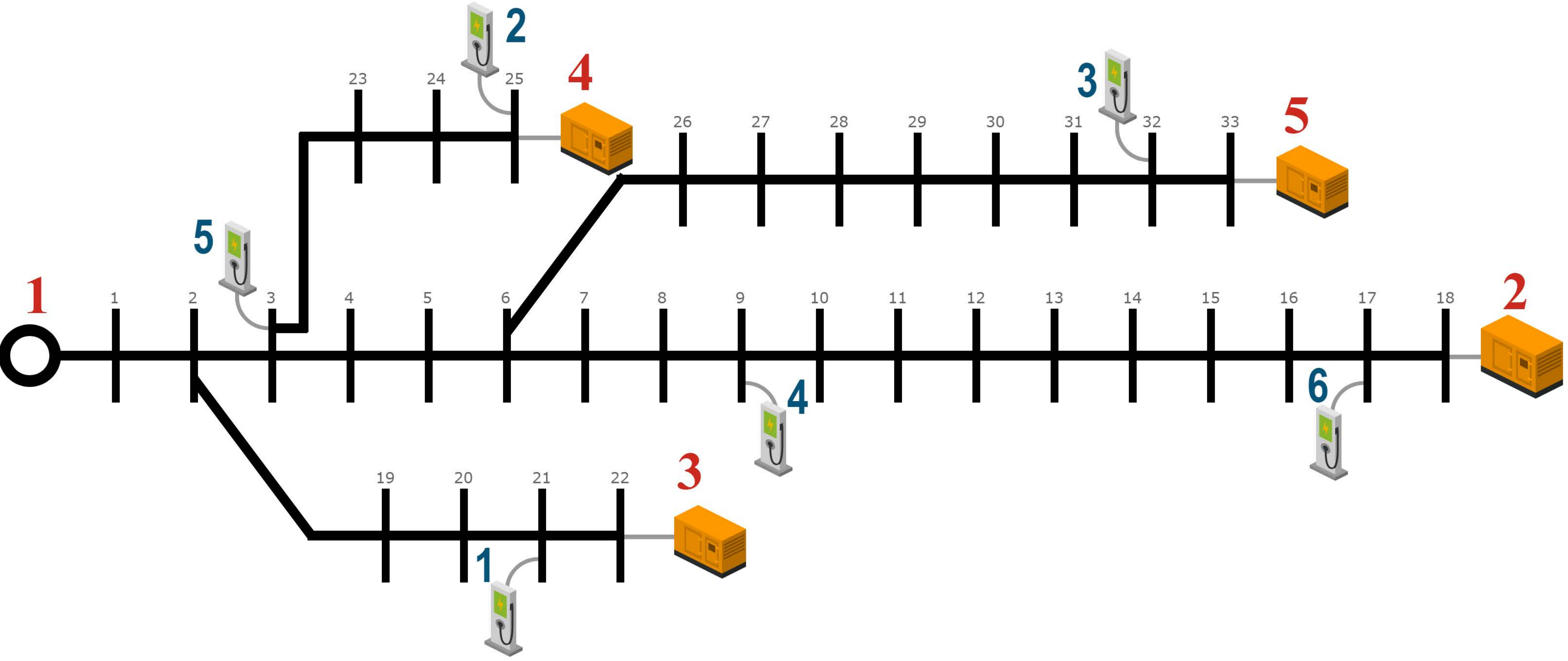}
	\caption{IEEE 33-bus distribution network, where CSs and DGs are located}
	\label{33_Bus}
\end{figure}
Detailed information on the distribution network {resistances and reactants} can be found in \cite{25627}, while the cost function and the generation limits of DGs are presented in Table \ref{generator_coeff}, except the parameters $\underline{p}_{h}^\mathrm{gen}$, $\underline{q}_{h}^\mathrm{gen}$ and ${{c}_{h}}$ which are set to 0. { We also assume that the substation bus is regulated and has a fixed voltage of 12.66 kV. Lastly, $\overline{v}_{j}$ and $\underline{v}_{j}$ are set to $193.93$ and $129.83$ at all nodes $j$.}
In Table \ref{lines}, we also provide the distribution network line limits. In addition, Table \ref{stations_limit} shows stations' coefficients. We choose these parameters to demonstrate how different DLMPs could arise in response to distribution line constraints and the heterogeneity of DG cost functions, such that they almost reflect the median electricity price in Savannah city \cite{eia}.
\begin{table}[!ht]
	\renewcommand{\arraystretch}{1.3}
	\caption{Cost function and constraint coefficients for DGs}
	\label{generator_coeff}
	\centering
	\begin{tabular}{| c | c | c | c | c | c | c |}
		\hline
		\bfseries Generation unit & \bfseries 1 & \bfseries 2 & \bfseries 3 & \bfseries 4 & \bfseries 5\\
		\hline
		$\overline{p}_{h}^\mathrm{gen} \text{MWh}$ & 5 & 2.3 & 1.5 & 2.2 & 1.6 \\
		\hline
			$\overline{q}_{h}^\mathrm{gen} \text{MVARh}$ & 5 & 2.3 & 1.5 & 2.2 & 1.6 \\
		\hline
		${{a}_{h}} (\$/ \text{MW}^2\text{h})$ & 0.35 & 0.2 & 0.4 & 0.3 & 0.25 \\
		\hline
		${{b}_{h}} (\$/ \text{MWh})$ & 76 & 45 & 88 & 66 & 56 \\
		\hline
	\end{tabular}
\end{table}
\begin{table}[!ht]
	\renewcommand{\arraystretch}{1.3}
	\caption{Distribution Network constraints coefficients for the distribution lines}
	\label{lines}
	\centering
	\begin{tabular}{| c | c | c | c | c | c | c |}
		\hline
		\bfseries Line & \bfseries 14 & \bfseries 20 & \bfseries 22 & \bfseries 30 \\
		\hline
		From bus & 14 & 20 & 3 & 30  \\
		\hline
		To bus & 15 & 21 & 23 & 31  \\
		\hline
		$S^\text{max}$ ($\text{kVAh}$) & 400 & 100 & 200 & 150  \\
		\hline
	\end{tabular}
\end{table}
\begin{table}[!ht]
	\renewcommand{\arraystretch}{1.1}
	\caption{Price and capacity coefficients for CSs}
	\label{stations_limit}
	\centering
	\begin{tabular}{| c | c | c | c | c | c | c |}
		\hline
		\bfseries Station & \bfseries 1 & \bfseries 2 & \bfseries 3 & \bfseries 4 & \bfseries 5 & \bfseries 6\\
		\hline
		${c_d^t}\left( \text{kWh}\right)$ & 1250 & 1000 & 1400 & 1450 & 1250 & 1350\\
		\hline
		$\zeta_d \left( \text{\$}\right)$ & 2 & 2 & 2 & 2 & 2 & 2\\
		\hline
		$\gamma_d \left( 1/h \right)$ & 2 & 2 & 2 & 2 & 2 & 2\\
		\hline
		$\phi_d$ & 12 & 14 & 8 & 10 & 12 & 8\\
		\hline
	\end{tabular}
\end{table}\\
First, we investigate a case in which there are 125 cars on the road with uniform distributions of electricity demand $q$ (kWh) $\sim U(20, 70)$ and monetary value of time $\omega_{i}(\$/\text{hr})\sim U(3.6, 14.4)$, as well as a uniform traffic distribution of $s_e(\frac{\text{vehicles}}{\text{hr}})\sim U(55, 150)$ on roads. Each roadway has a maximum vehicle capacity of $160$. We assumed that EVs were dispersed uniformly over the network. Additionally, just for simplicity, we assumed that each EV $i$ parameters $\alpha_{i}$ and $\beta_{i}$ are equal $0.5 \$ $ and likes to take the shortest route to the nearest station, which has effects on parameters $\tilde{r}$‌ and $\tilde{t}$.\\
The first set of analyses investigates the convergence of the algorithm. Fig. \ref{Gen_s1} shows the convergence of DGs' output power. We notice that DG 5, which has a lower operating cost than DGs 1, 3, and 4, and is located close to station 3, has reached its generation capacity due to the high demand of station 3.
\begin{figure}[!ht]
	\centering
	\includegraphics[width=1\linewidth]{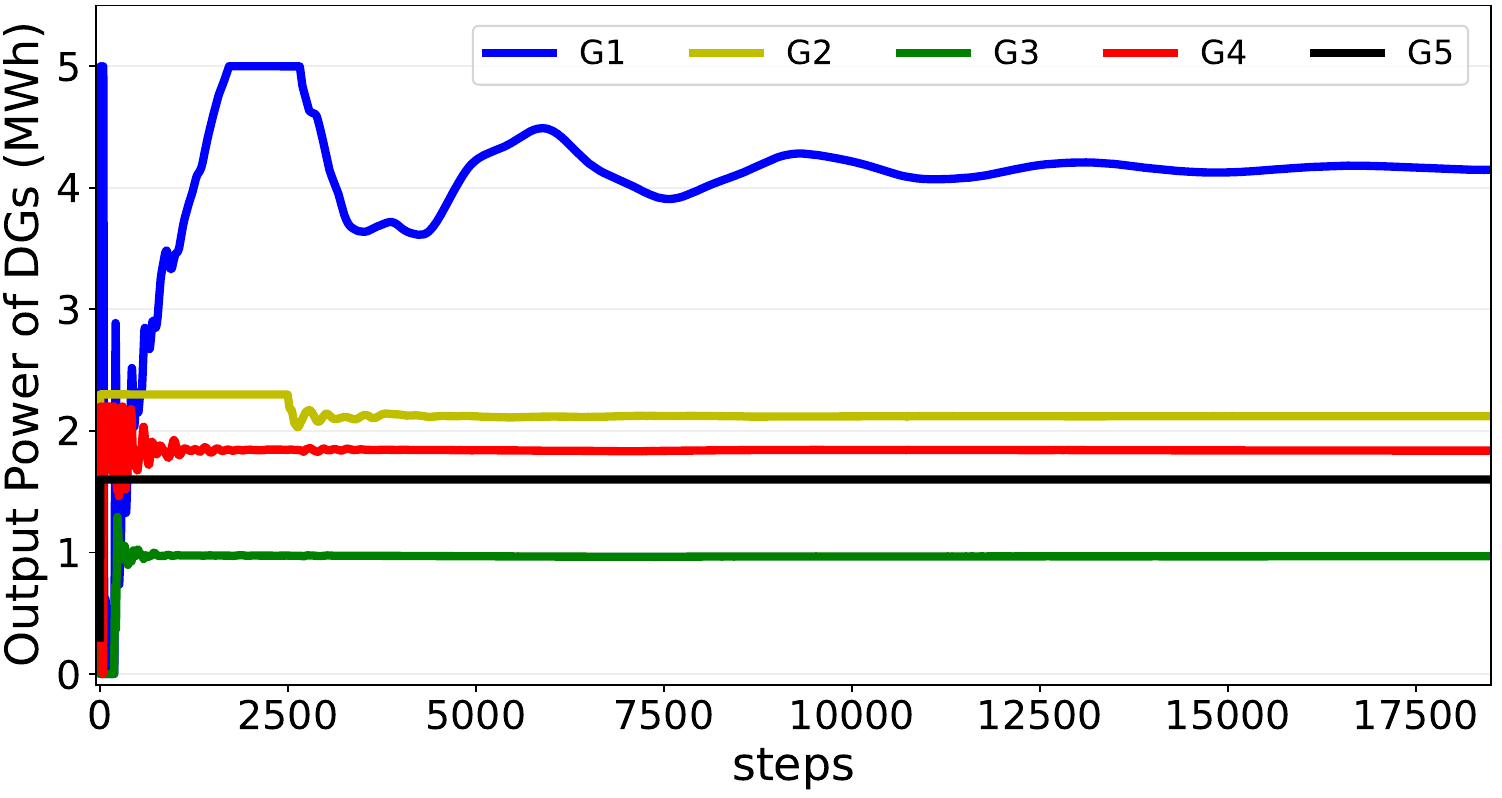}
	\caption{Active output powers of DGs, which depend on their costs, capacities, lines' constraints, and nearby demand}
	\label{Gen_s1}
\end{figure}
In Fig. \ref{LMP_s1}, we see the convergence of DLMPs at the buses where charging stations are located. The differences in DLMPs across locations are due to different costs associated with the generation units and the fact that all four distribution lines presented in Table \ref{lines} are operating at full capacity.
\begin{figure}[!ht]
	\centering
	\includegraphics[width=1\linewidth]{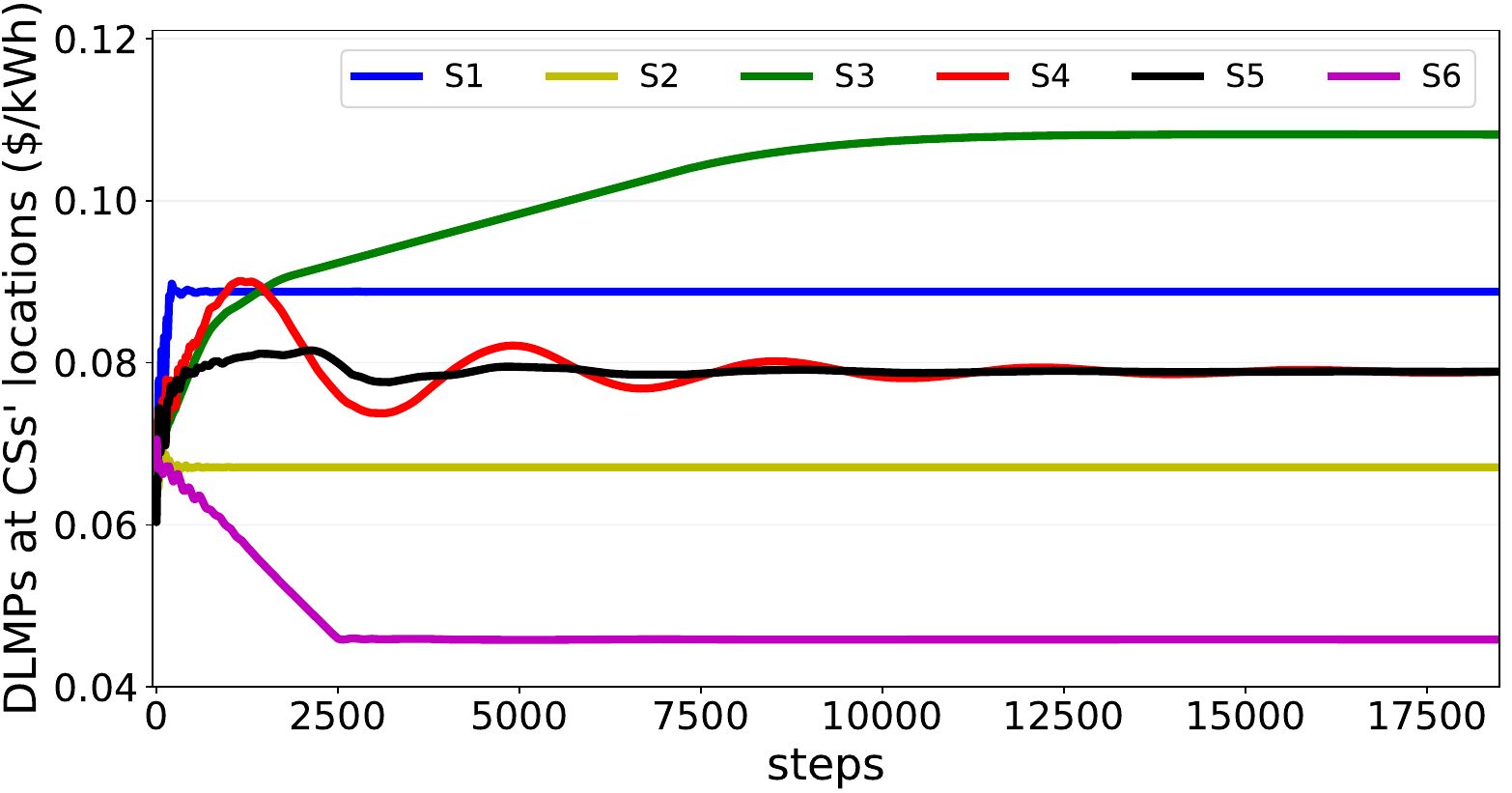}
	\caption{DLMPs at the CSs' locations; in areas close to DGs with higher costs and demand, prices are higher.}
	\label{LMP_s1}
\end{figure}
Despite the fact that the generation unit price is cheaper in the area surrounding station 3, its DLMP is the highest. This is due to the fact that user preferences, and consequently energy demand, exceed the capacity of the power network to deliver them. Therefore, the DSO must increase the DLMPs to a level that incentivizes EVs to use stations with lower operational costs. In addition, Fig. \ref{Demand_s1} depicts the convergence of aggregated demand at stations, and Fig. \ref{Surcharge_s1} depicts the corresponding tolls for each charging station. 
\begin{figure}[!ht]
	\centering
	\includegraphics[width=1\linewidth]{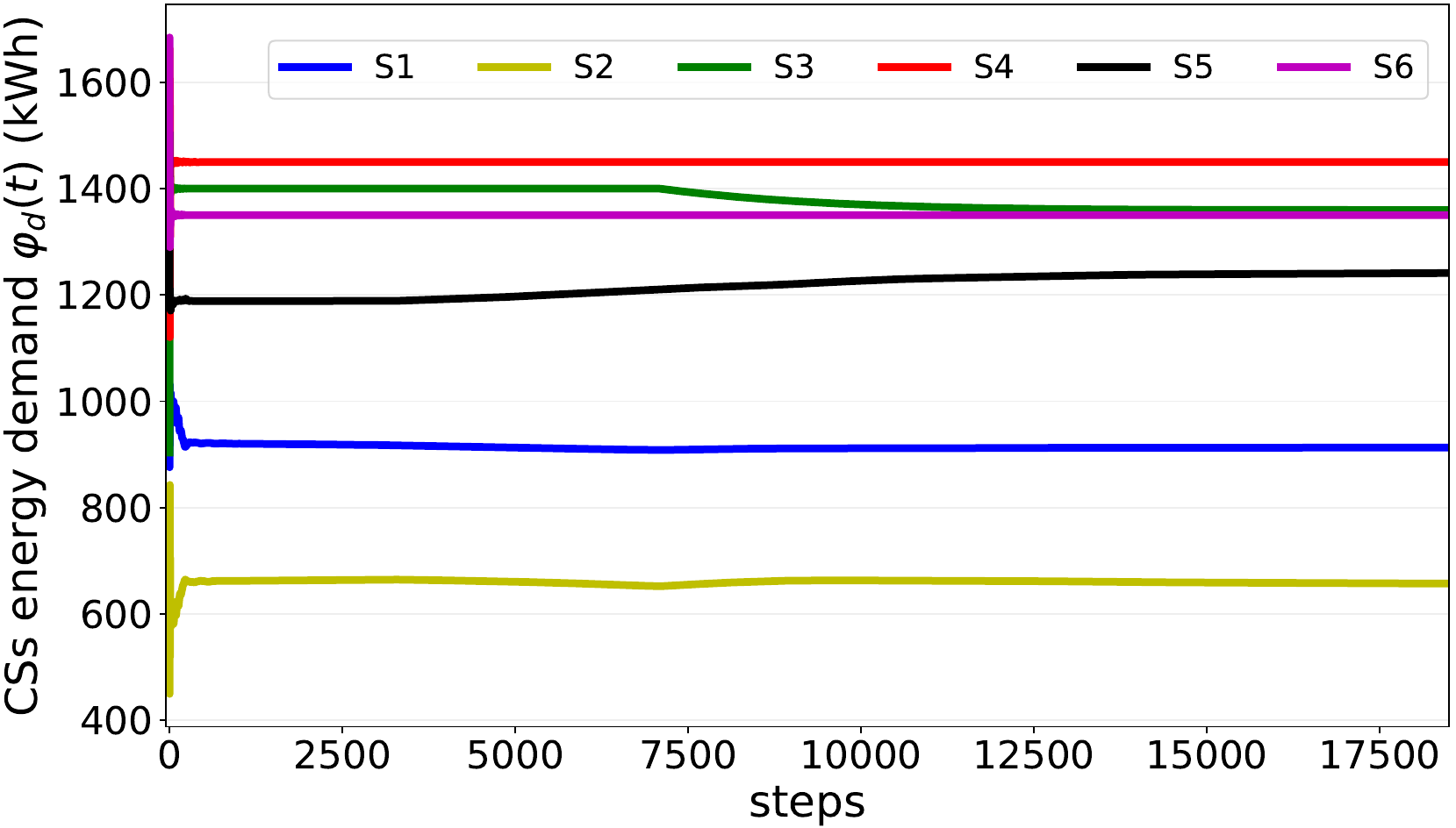}
	\caption{Demand for energy at CSs; capacity limits are met}
	\label{Demand_s1}
\end{figure}
\begin{figure}[!ht]
	\centering
	\includegraphics[width=1\linewidth]{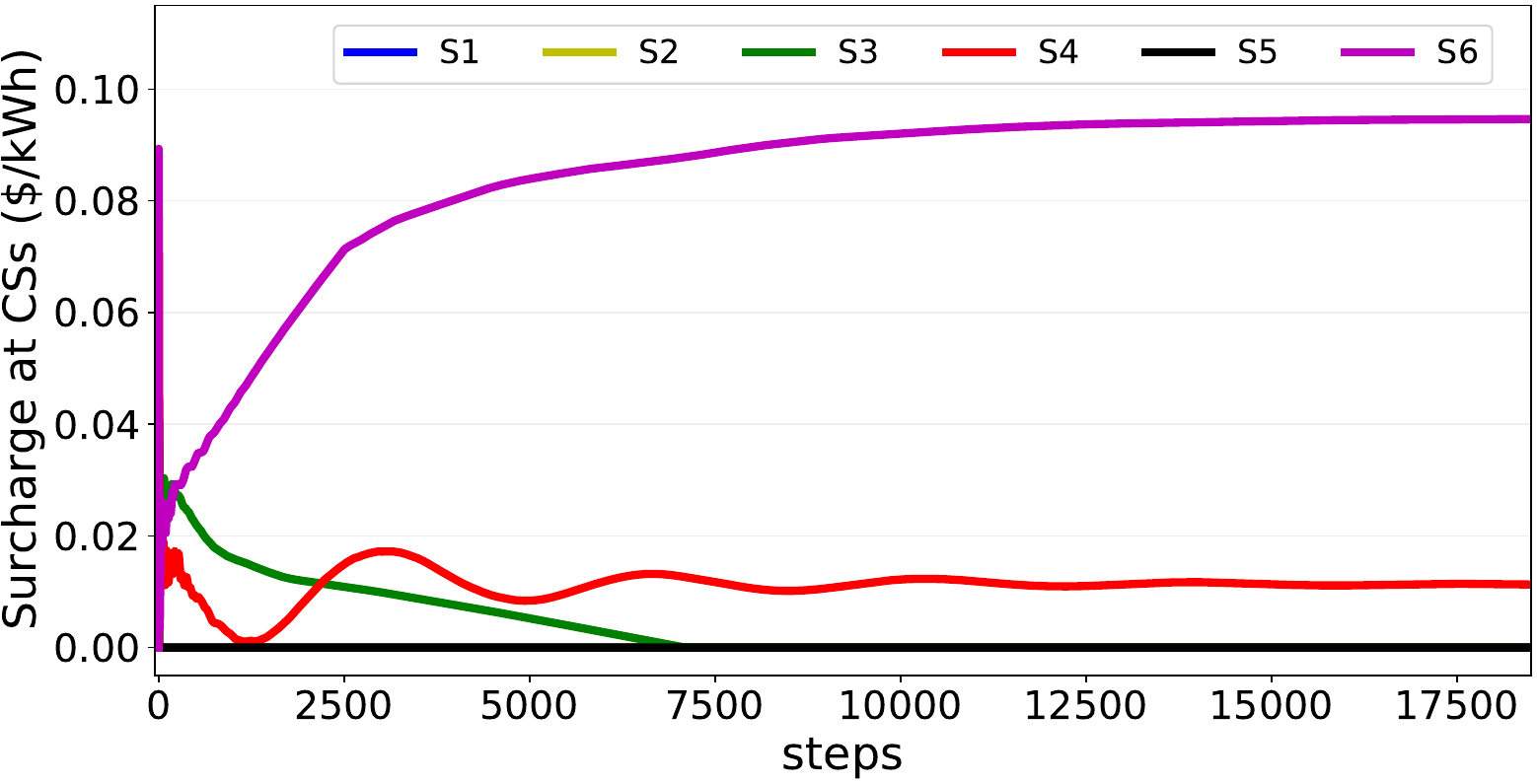}
	\caption{Surcharges imposed by CSs to enforce capacity limits}
	\label{Surcharge_s1}
\end{figure}
In addition, we initially see how stations 3, 4, and 6 control their demand capacity by applying local surcharges. Nonetheless, the DSO deems this management insufficient. By increasing DLMP by the DSO at station 3, the demand at that station falls below its capacity; consequently, the local surcharge disappears. {In addition, Fig. \ref{bus} shows the voltage magnitudes of buses across the distribution network. We can see that the lowest voltage magnitude is at bus 9, where charging station 4, with high demand, is located. Another reason is that there is no generator nearby this bus.\\ Simultaneously, the TNO manages the roads by charging users. For example, the southern road of station 4 reached its capacity limits, and the TNO charges 3.44\$ for using that roadway. Consequently, users change their route and use the eastern road to reach station 4 instead of the southern road. All computational simulations are performed on Google Colab with 12GB memory. At each iteration, the average computation time of each EV, the DSO, and the TNO are 0.074, 0.17, and 0.004 seconds, respectively.}
\begin{figure}[!ht]
	\centering
	\includegraphics[width=1\linewidth]{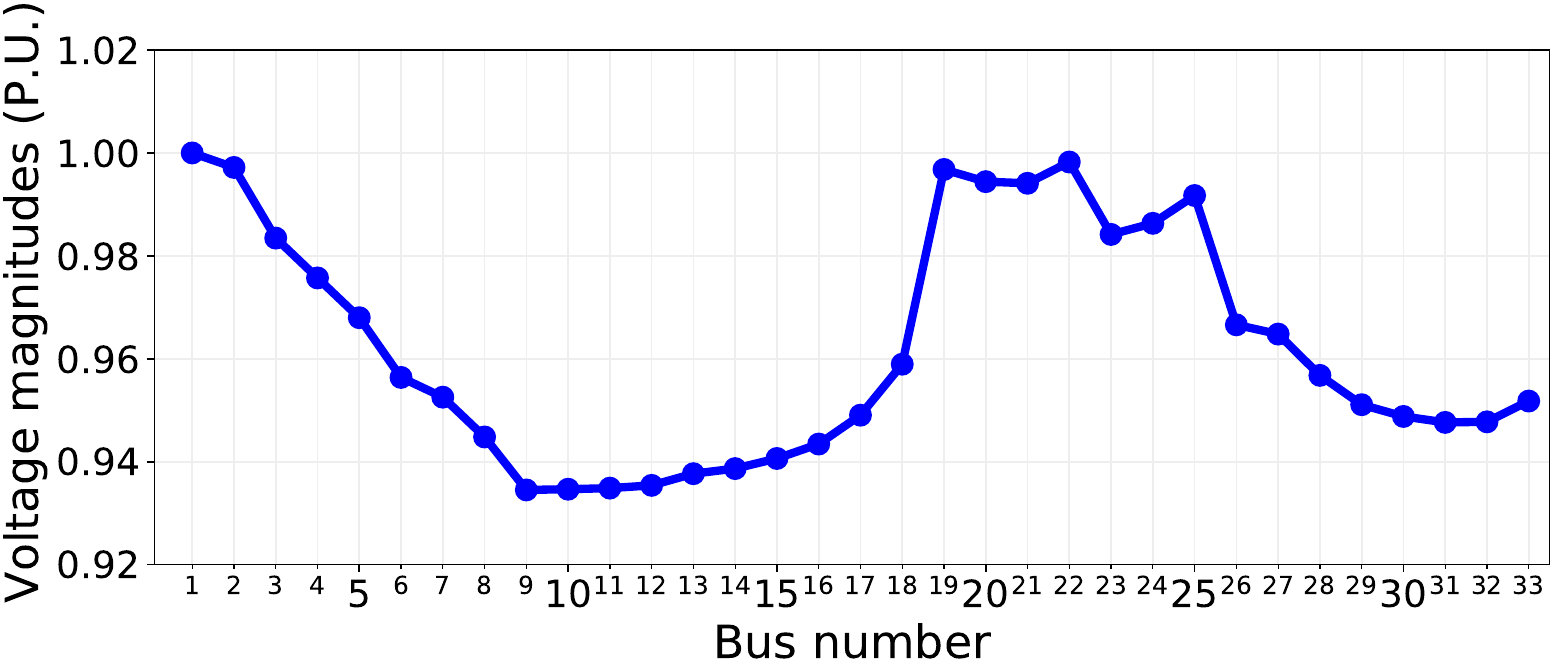}
	\caption{Voltage magnitudes of buses throughout the network}
	\label{bus}
\end{figure}
\\
Afterward, we investigate the effects of various elements on the DSO operation. We disregard the limited coupling constraints of stations and roadways for this purpose.
First, we analyze the effect of the number of EVs (and thus the total electricity demand) on the DLMPs.
As shown in Fig. \ref{Compare}-a, while demand is low, station 3 has a lower DLMP than stations 1, 2, 4, and 5; nevertheless, increasing the number of EVs raises the DLMP, since in this case stronger persuasion is needed to change the EVs' preferred decision. Also, we can observe that DLMP at station 6 is increasing since DG 2 and distribution line 14 have reached their limits, and the DSO is unable to supply adequate electricity to that station.\\
We study the impact of a personal parameter, known as the monetary value of time, on the DSO decision-making. In Fig. \ref{Compare}-b, as the mean value of the time value of money increases, so do the DLMP on popular location like CS 4 because EVs require more persuasion to change their decisions.
\begin{figure}[!ht]
	\centering
	\includegraphics[width=1\linewidth]{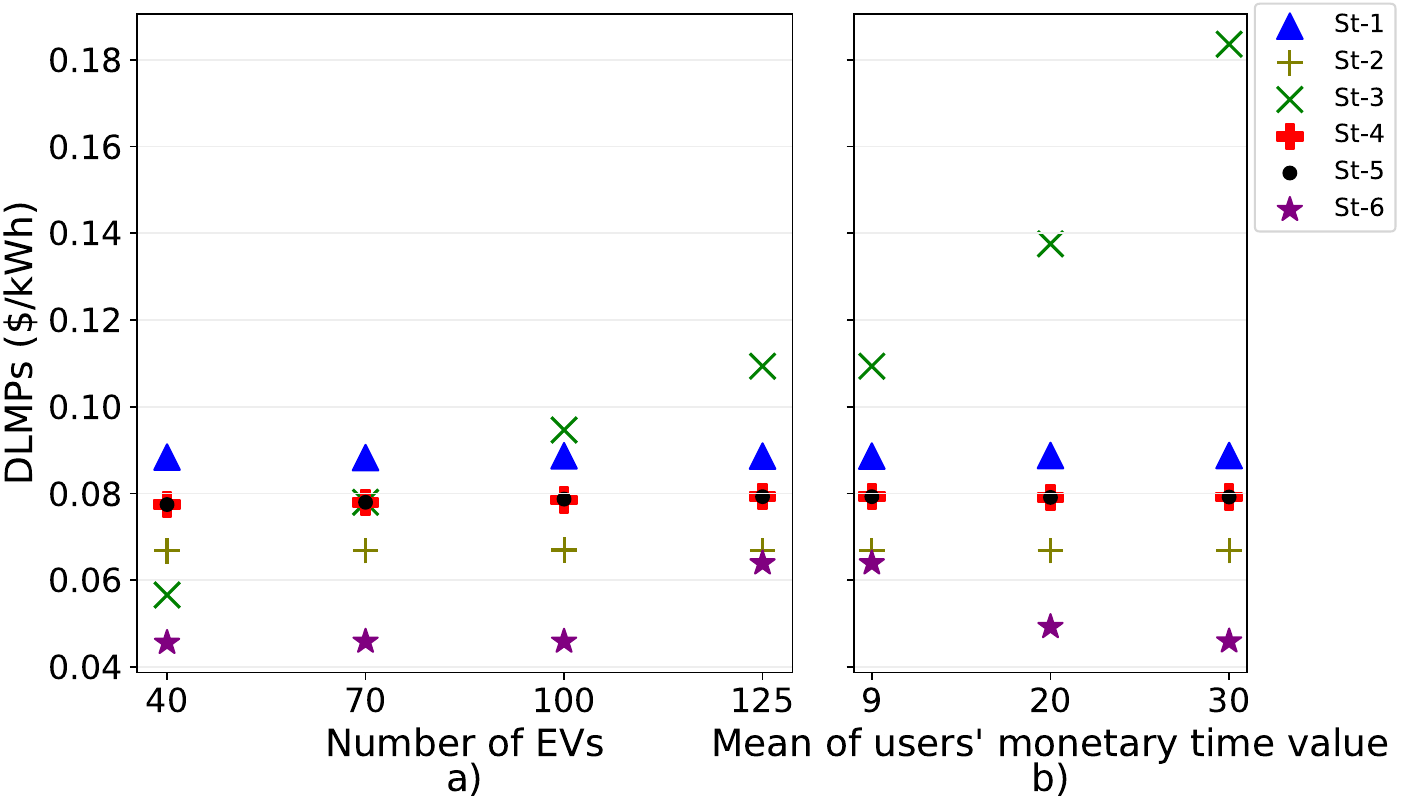}
	\caption{Effects of users' parameters on the DSO's decision}
	\label{Compare}
\end{figure}
\section{Conclusion and Future works}\label{Conclusion}
With the use of a generalized aggregative game, we modeled and evaluated the interactions of heterogeneous EVs, the DSO, CSO, and the TNO. We presented a decentralized learning technique based on real-time DLMP by the DSO and pricing and tolling by CSOs and the TNO to satisfy infrastructure restrictions.
The decentralized learning technique is scalable and secure. It also leads to a generalized Wardrop equilibrium of the holistic model. We study the Savannah city model and the IEEE 33-bus distribution network numerically. We also looked into the influence of station characteristics on pricing and demand. {This work can be extended in several directions, including mechanism design for better system-level performance, considering dynamic and time-varying models for more precise modeling of the problem, exploration of multi-agent reinforcement learning for solving the problem, and faster equilibrium seeking algorithms for practical implementation of the proposed scheme.}
\bibliographystyle{IEEEtraN}
\bibliography{Ref}

\begin{thebibliography}{10}
\providecommand{\url}[1]{#1}
\csname url@samestyle\endcsname
\providecommand{\newblock}{\relax}
\providecommand{\bibinfo}[2]{#2}
\providecommand{\BIBentrySTDinterwordspacing}{\spaceskip=0pt\relax}
\providecommand{\BIBentryALTinterwordstretchfactor}{4}
\providecommand{\BIBentryALTinterwordspacing}{\spaceskip=\fontdimen2\font plus
\BIBentryALTinterwordstretchfactor\fontdimen3\font minus
  \fontdimen4\font\relax}
\providecommand{\BIBforeignlanguage}[2]{{%
\expandafter\ifx\csname l@#1\endcsname\relax
\typeout{** WARNING: IEEEtran.bst: No hyphenation pattern has been}%
\typeout{** loaded for the language `#1'. Using the pattern for}%
\typeout{** the default language instead.}%
\else
\language=\csname l@#1\endcsname
\fi
#2}}
\providecommand{\BIBdecl}{\relax}
\BIBdecl

\bibitem{bibra2021global}
\BIBentryALTinterwordspacing
E.~M. Bibra, E.~Connelly, M.~Gorner, C.~Lowans, L.~Paoli, J.~Tattini, and
  J.~Teter, \emph{Global EV Outlook 2021: Accelerating Ambitions Despite the
  Pandemic}.\hskip 1em plus 0.5em minus 0.4em\relax IEA, 2021. [Online].
  Available: \url{https://www.iea.org/reports/global-ev-outlook-2021}
\BIBentrySTDinterwordspacing

\bibitem{davis2018net}
S.~J. Davis, N.~S. Lewis, M.~Shaner, S.~Aggarwal, D.~Arent, I.~L. Azevedo,
  S.~M. Benson, T.~Bradley, J.~Brouwer, Y.-M. Chiang \emph{et~al.}, ``Net-zero
  emissions energy systems,'' \emph{Science}, vol. 360, no. 6396, p. eaas9793,
  2018.

\bibitem{muratori2018impact}
M.~Muratori, ``Impact of uncoordinated plug-in electric vehicle charging on
  residential power demand,'' \emph{Nature Energy}, vol.~3, no.~3, pp.
  193--201, 2018.

\bibitem{18}
B.~G. Bakhshayesh and H.~Kebriaei, ``Decentralized equilibrium seeking of joint
  routing and destination planning of electric vehicles: A constrained
  aggregative game approach,'' \emph{IEEE Transactions on Intelligent
  Transportation Systems}, pp. 1--10, 2021.

\bibitem{ratliff2019perspective}
L.~J. Ratliff, R.~Dong, S.~Sekar, and T.~Fiez, ``A perspective on incentive
  design: Challenges and opportunities,'' \emph{Annual Review of Control,
  Robotics, and Autonomous Systems}, vol.~2, pp. 305--338, 2019.

\bibitem{3}
M.~Alizadeh, H.-T. Wai, M.~Chowdhury, A.~Goldsmith, A.~Scaglione, and
  T.~Javidi, ``Optimal pricing to manage electric vehicles in coupled power and
  transportation networks,'' \emph{IEEE Transactions on control of network
  systems}, vol.~4, no.~4, pp. 863--875, 2016.

\bibitem{5}
Y.~Sun, Z.~Chen, Z.~Li, W.~Tian, and M.~Shahidehpour, ``Ev charging schedule in
  coupled constrained networks of transportation and power system,'' \emph{IEEE
  Transactions on Smart Grid}, vol.~10, no.~5, pp. 4706--4716, 2018.

\bibitem{10}
F.~Rossi, R.~Iglesias, M.~Alizadeh, and M.~Pavone, ``On the interaction between
  autonomous mobility-on-demand systems and the power network: Models and
  coordination algorithms,'' \emph{IEEE Transactions on Control of Network
  Systems}, vol.~7, no.~1, pp. 384--397, 2020.

\bibitem{11}
A.~Estandia, M.~Schiffer, F.~Rossi, J.~Luke, E.~C. Kara, R.~Rajagopal, and
  M.~Pavone, ``On the interaction between autonomous mobility on demand systems
  and power distribution networks—an optimal power flow approach,''
  \emph{IEEE Transactions on Control of Network Systems}, vol.~8, no.~3, pp.
  1163--1176, 2021.

\bibitem{1}
F.~He, Y.~Yin, J.~Wang, and Y.~Yang, ``Sustainability si: optimal prices of
  electricity at public charging stations for plug-in electric vehicles,''
  \emph{Networks and Spatial Economics}, vol.~16, no.~1, pp. 131--154, 2016.

\bibitem{2}
W.~Wei, L.~Wu, J.~Wang, and S.~Mei, ``Network equilibrium of coupled
  transportation and power distribution systems,'' \emph{IEEE Transactions on
  Smart Grid}, vol.~9, no.~6, pp. 6764--6779, 2017.

\bibitem{6}
W.~Wei, S.~Mei, L.~Wu, M.~Shahidehpour, and Y.~Fang, ``Optimal traffic-power
  flow in urban electrified transportation networks,'' \emph{IEEE Transactions
  on Smart Grid}, vol.~8, no.~1, pp. 84--95, 2016.

\bibitem{7}
T.~Qian, C.~Shao, X.~Li, X.~Wang, and M.~Shahidehpour, ``Enhanced coordinated
  operations of electric power and transportation networks via ev charging
  services,'' \emph{IEEE Transactions on Smart Grid}, vol.~11, no.~4, pp.
  3019--3030, 2020.

\bibitem{12}
B.~Sohet, Y.~Hayel, O.~Beaude, and A.~Jeandin, ``Hierarchical coupled
  driving-and-charging model of electric vehicles, stations and grid
  operators,'' \emph{IEEE Transactions on Smart Grid}, pp. 1--1, 2021.

\bibitem{13}
W.~Wei, W.~Danman, W.~Qiuwei, M.~Shafie-Khah, and J.~P. Catalao,
  ``Interdependence between transportation system and power distribution
  system: A comprehensive review on models and applications,'' \emph{Journal of
  Modern Power Systems and Clean Energy}, vol.~7, no.~3, pp. 433--448, 2019.

\bibitem{zardini2022analysis}
G.~Zardini, N.~Lanzetti, M.~Pavone, and E.~Frazzoli, ``Analysis and control of
  autonomous mobility-on-demand systems,'' \emph{Annual Review of Control,
  Robotics, and Autonomous Systems}, vol.~5, pp. 633--658, 2022.

\bibitem{8}
Q.~Chen, F.~Wang, B.-M. Hodge, J.~Zhang, Z.~Li, M.~Shafie-Khah, and J.~P.
  Catal{\~a}o, ``Dynamic price vector formation model-based automatic demand
  response strategy for pv-assisted ev charging stations,'' \emph{IEEE
  Transactions on Smart Grid}, vol.~8, no.~6, pp. 2903--2915, 2017.

\bibitem{9}
A.~Moradipari, N.~Tucker, and M.~Alizadeh, ``Mobility-aware electric vehicle
  fast charging load models with geographical price variations,'' \emph{IEEE
  Transactions on Transportation Electrification}, vol.~7, no.~2, pp. 554--565,
  2021.

\bibitem{tucker2019online}
N.~Tucker, B.~Turan, and M.~Alizadeh, ``Online charge scheduling for electric
  vehicles in autonomous mobility on demand fleets,'' in \emph{2019 IEEE
  Intelligent Transportation Systems Conference (ITSC)}.\hskip 1em plus 0.5em
  minus 0.4em\relax IEEE, 2019, pp. 226--231.

\bibitem{9390363}
K.-F. Chu, A.~Y.~S. Lam, and V.~O.~K. Li, ``Joint rebalancing and
  vehicle-to-grid coordination for autonomous vehicle public transportation
  system,'' \emph{IEEE Transactions on Intelligent Transportation Systems},
  vol.~23, no.~7, pp. 7156--7169, 2022.

\bibitem{113}
N.~I. Nimalsiri, C.~P. Mediwaththe, E.~L. Ratnam, M.~Shaw, D.~B. Smith, and
  S.~K. Halgamuge, ``A survey of algorithms for distributed charging control of
  electric vehicles in smart grid,'' \emph{IEEE Transactions on Intelligent
  Transportation Systems}, vol.~21, no.~11, pp. 4497--4515, 2020.

\bibitem{4}
R.~Cole, Y.~Dodis, and T.~Roughgarden, ``Pricing network edges for
  heterogeneous selfish users,'' in \emph{Proceedings of the thirty-fifth
  annual ACM symposium on Theory of computing}, 2003, pp. 521--530.

\bibitem{Bureau}
{Bureau of Public Roads}, \emph{Traffic assignment manual}.\hskip 1em plus
  0.5em minus 0.4em\relax U.S. Dept. of Commerce, Urban Planning Division,
  Tech. Rep., 1964.

\bibitem{reid2019operations}
R.~D. Reid and N.~R. Sanders, \emph{Operations management: an integrated
  approach}.\hskip 1em plus 0.5em minus 0.4em\relax John Wiley \& Sons, 2019.

\bibitem{baran1989optimal}
M.~E. Baran and F.~F. Wu, ``Optimal capacitor placement on radial distribution
  systems,'' \emph{IEEE Transactions on power Delivery}, vol.~4, no.~1, pp.
  725--734, 1989.

\bibitem{zhu2015fast}
H.~Zhu and H.~J. Liu, ``Fast local voltage control under limited reactive
  power: Optimality and stability analysis,'' \emph{IEEE Transactions on Power
  Systems}, vol.~31, no.~5, pp. 3794--3803, 2015.

\bibitem{wardrop1952road}
J.~G. Wardrop, ``Road paper. some theoretical aspects of road traffic
  research.'' \emph{Proceedings of the institution of civil engineers}, vol.~1,
  no.~3, pp. 325--362, 1952.

\bibitem{14}
G.~Belgioioso and S.~Grammatico, ``Semi-decentralized generalized nash
  equilibrium seeking in monotone aggregative games,'' \emph{IEEE Transactions
  on Automatic Control}, pp. 1--1, 2021.

\bibitem{16}
D.~Paccagnan, B.~Gentile, F.~Parise, M.~Kamgarpour, and J.~Lygeros, ``Nash and
  wardrop equilibria in aggregative games with coupling constraints,''
  \emph{IEEE Transactions on Automatic Control}, vol.~64, no.~4, pp.
  1373--1388, 2018.

\bibitem{Palomar2009}
D.~P. Palomar and Y.~C. Eldar, ``{Convex optimization in signal processing and
  communications},'' \emph{Convex Optimization in Signal Processing and
  Communications}, vol. 9780521762, no.~2, pp. 1--498, 2009.

\bibitem{114}
F.~Faccchinei and J.~Pang, ``Finite-dimensional variational inequalities and
  complementarity problem,'' 2003.

\bibitem{auslender2000lagrangian}
A.~Auslender and M.~Teboulle, ``Lagrangian duality and related multiplier
  methods for variational inequality problems,'' \emph{SIAM Journal on
  Optimization}, vol.~10, no.~4, pp. 1097--1115, 2000.

\bibitem{15}
Y.~Malitsky and M.~K. Tam, ``A forward-backward splitting method for monotone
  inclusions without cocoercivity,'' \emph{SIAM Journal on Optimization},
  vol.~30, no.~2, pp. 1451--1472, 2020.

\bibitem{115}
D.~Calderone, E.~Mazumdar, L.~J. Ratliff, and S.~S. Sastry, ``Understanding the
  impact of parking on urban mobility via routing games on queue-flow
  networks,'' in \emph{2016 IEEE 55th Conference on Decision and Control
  (CDC)}.\hskip 1em plus 0.5em minus 0.4em\relax IEEE, 2016, pp. 7605--7610.

\bibitem{25627}
M.~Baran and F.~Wu, ``Network reconfiguration in distribution systems for loss
  reduction and load balancing,'' \emph{IEEE Transactions on Power Delivery},
  vol.~4, no.~2, pp. 1401--1407, 1989.

\bibitem{eia}
\BIBentryALTinterwordspacing
Average price of electricity to ultimate customers by end-use sector. [Online].
  Available: \url{https://www.eia.gov/electricity/monthly/}
\BIBentrySTDinterwordspacing

\end{thebibliography}
\end{document}